\documentclass[%
 amsmath,amssymb,
 reprint,%
superscriptaddress,
]{revtex4-2}

\usepackage{graphicx}
\usepackage{dcolumn}
\usepackage{bm}
\usepackage[hidelinks]{hyperref}

\usepackage[utf8]{inputenc}
\usepackage[T1]{fontenc}
\usepackage{mathptmx}
\usepackage{etoolbox}

\begin{document}


\title[]{Circle fit optimization for resonator quality factor measurements: point redistribution for maximal accuracy}
\author{Paul G. Baity}
\author{Connor Maclean}
\author{Valentino Seferai}
\author{Joe Bronstein}
\affiliation{James Watt School of Engineering and Centre for Quantum Technology, University of Glasgow, Glasgow, G12 8QQ, UK}

\author{Yi Shu}
\author{Tania Hemakumara}
\affiliation{Oxford Instruments Plasma Technology, North End, Yatton, Bristol, BS49 4AP, UK}
\author{Martin Weides}
\affiliation{James Watt School of Engineering and Centre for Quantum Technology, University of Glasgow, Glasgow, G12 8QQ, UK}

\date{\today}

\begin{abstract}
The control of material loss mechanisms is playing an increasingly important role for improving coherence times of superconducting quantum devices. Such material losses can be characterized through the measurement of planar superconducting resonators, which reflect losses through the resonance’s quality factor $Q_l$. The resonance quality factor consists of both internal (material) losses as well as coupling losses when resonance photons escape back into the measurement circuit. The combined losses are then described as $Q_l^{-1} = \mathrm{Re}\{Q_c^{-1}\} + Q_i^{-1}$, where $Q_c$ and $Q_i$ reflect the coupling and internal quality factors of the resonator, respectively. To separate the relative contributions of $Q_i$ and $Q_c$ to $Q_l$, diameter-correcting circle fits use algebraic or geometric means to fit the resonance signal on the complex plane. However, such circle fits can produce varied results, so to address this issue, we use a combination of simulation and experiment to determine the reliability of a fitting algorithm across a wide range of quality factor values from $Q_i\ll Q_c$ to $Q_c\ll Q_i$. In addition, we develop a novel measurement protocol that can not only reduce fitting errors by factors $\gtrsim 2$ but also mitigates the influence of the measurement background on the fit results. This technique can be generalized for other resonance systems beyond superconducting resonators.

\end{abstract}

\maketitle
\section{Introduction}
Superconducting quantum technologies are the current leading platform for the implementation of scaled quantum computing, and large strides are being made to not only scale up superconducting circuits but also improve their performance by mitigating sources of error \cite{Place2021,Murray2021,Wang2022,Iaia2022,Verjauw2022}.
Coplanar waveguide (CPW) resonators are one of the key components for superconducting-based quantum computing platforms and are used for qubit readout and coupling \cite{Krantz2019,Rossi2021}. However, resonators can also be used to probe \cite{Megrant2012,Bruno2015,Dunsworth2017,Henriques2019,McRae2020,deGraaf2020,Altoe2022} the loss mechanisms that limit qubit coherence times in quantum computing processors. On the other hand, there are several methods used to characterize resonance signals \cite{Petersan1998,Mcrae2022}, and results can vary depending on the method used to characterize the resonance signal. However, a fair comparison between not only the performance of these various resonance fitting models but also the results across the broad spectrum of research that applies these models requires uncertainty quantification for these models and tests of their reliability  to accurately apply the model to experimental data. 

One of the more popular resonance fitting methods is the diameter corrected circle fit  \cite{Khalil2012,Megrant2012,Probst2015,Rieger2023}. This method uses either algebraic or geometric methods to fit the circular projection of the resonance signal on the complex plane to extract the resonance properties, such as the internal quality factor $Q_i$, which characterizes material losses. If $Q_l$ is the total resonance efficiency (i.e. quality factor) and $Q_c$ is the coupling efficiency to the measurement line, then the circle diameter is $d=Q_l/|Q_c|$. When $Q_l$ is determined from an independent method, such as a geometric fit of the resonance phase signal $\theta$, then $d$ can be used to distinguish $Q_c$ from the material loss efficiency factor $Q_i$ through the relationship $Q_l^{-1} = \mathrm{Re}\{Q_c^{-1}\} + Q_i^{-1}$. 

\begin{figure}
    \includegraphics[scale = 0.45]{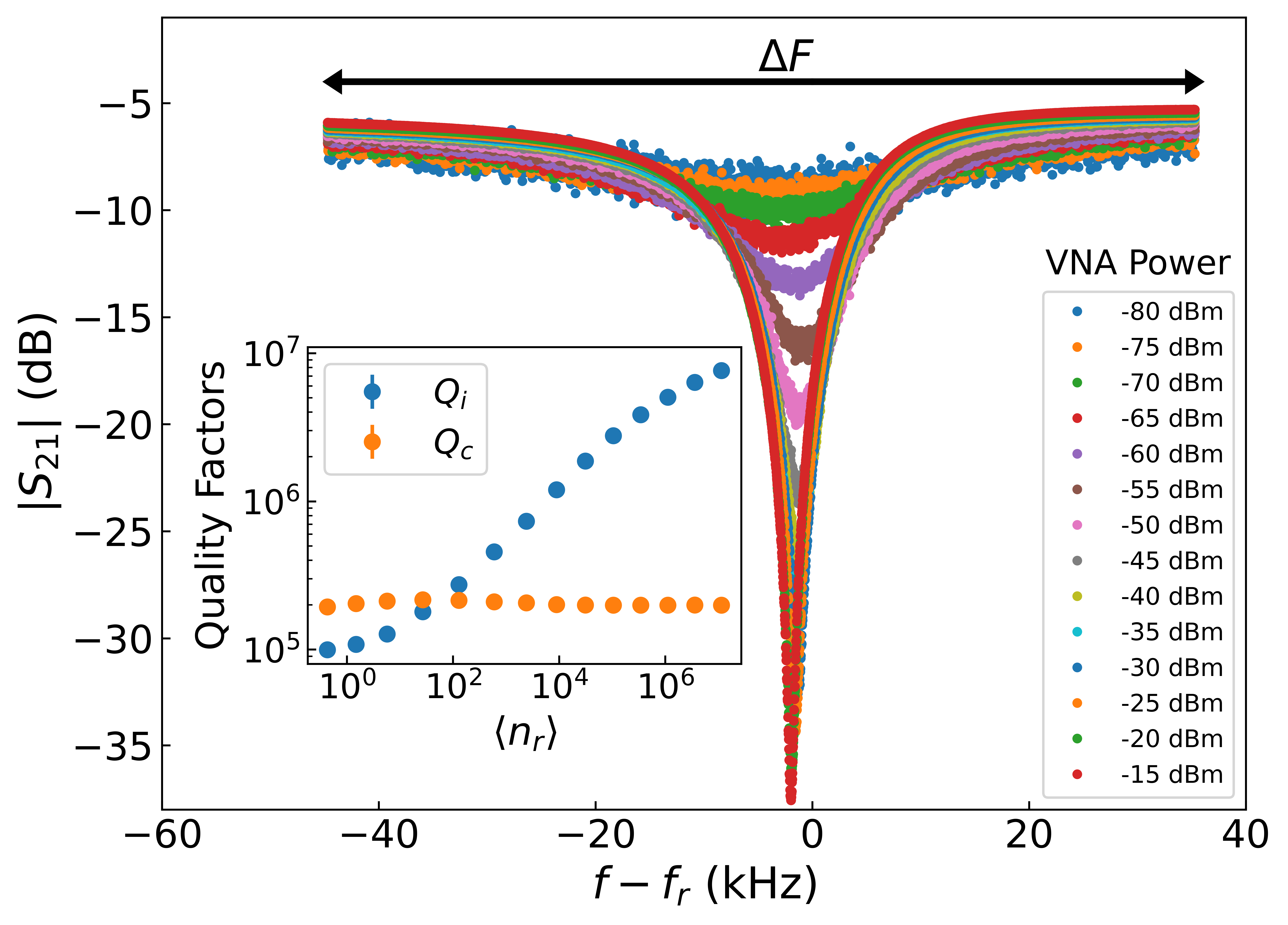}
    \caption{Measurement of a superconducting CPW microwave resonator fabricated on an ALD NbN film. Each curve is measured with $N = 4001$ points distributed linearly across the measurement bandwidth $\Delta F$ and with a specified VNA output power. A power-dependent trace averaging protocol was used as described in Appendix B. VNA signals were attenuated by an approximate -73 dB before propagating into the chip. As on-chip power ($P_{chip}$) increases, the resonance grows sharper as the linewidth decreases, signifying an increase in the loaded quality factor. (inset) The loaded quality factor can be devolved into internal and coupling quality factors, $Q_i$ and $Q_c$ respectively, using a diameter correcting circle fit \cite{Probst2015}. As the photon number $\langle n_r \rangle$ decreases, $Q_i$ is also observed to decrease due to the presence of TLS losses. $\langle n_r \rangle$ is calculated using Eq.~\ref{eq:nr} derived in Appendix E.}
    \label{fig:powerdep}
\end{figure}

However, resonances can be difficult to measure or fit in extreme limits when $Q_c$ and $Q_i$ differ by orders of magnitude. This is further complicated by the measurement setup, which can introduce systematic errors into the analysis \cite{Probst2015,Rieger2023}. Fig.~\ref{fig:powerdep} shows a measurement of a CPW resonator in a range of signal power levels. As power and the corresponding resonance photon number $\langle n_r \rangle$ decrease, the resonance signal becomes less sharp as the $Q_i$ decreases due to increased two-level system (TLS) losses \cite{Macha2010,Sage2011,Vissers2012,Bruno2015,Dunsworth2017,Calusine2018,McRae2018,McRae2020,Altoe2022}. At the lowest powers, the resonance signal is almost lost into the microwave background signal, which places a limitation on the extraction of $Q_i$ at low powers. A method for bypassing this issue is to use slightly or moderately overcoupled resonators to facilitate low-power measurements. However, when $Q_c$ is relatively small, the total contribution of $Q_i$ to the loaded quality factor also becomes increasingly small. To better understand the limitations imposed by design on $Q_i$ determination, we perform a benchmark on the fitting algorithm of Probst et al. \cite{Probst2015} and an analysis of the model uncertainties to detail how fitting errors behave under extreme conditions when $Q_c \gg Q_i$. We will also perform tests for potential sources of fit bias and will introduce an alternative measurement point distribution to mitigate such biases. The analytical techniques used here can also be applied to other fitting methods, which is an important first step toward reconciling any differences in results that they may return \cite{Mcrae2022}.

\section{Modeling}
For a notch-type resonator coupled to a transmission line, the frequency-dependent transmission $\tilde{S}_{21}$ is given by \cite{Probst2015, qkit}
\begin{equation}
\tilde{S}_{21} = a e^{i\alpha-2\pi i f\tau} \bigg[ 1 - \frac{(Q_l/|Q_c|)e^{i\phi}}{1 + 2 iQ_l(f/f_r-1)} \bigg],
\label{resEq}
\end{equation}
where $f$ is the signal frequency, $f_r$ is the resonance frequency, and $\phi$ is the phase shift of resonance due to impedance mismatch with the transmission line \cite{Khalil2012}. The prefactor $a e^{i\alpha-2\pi i f\tau}$ models the effects of the microwave background environment, where $a$, $\alpha$, and $\tau$ are the background transmission amplitude, phase offset, and microwave delay, all of which arise from the specifics of the measurement setup. To assist the reader for this and subsequent sections, we have included a table of variables and their definitions in Appendix A. An artificial resonance signal can be generated using Eq.~\ref{resEq} and randomly generated noise. In practice, this microwave background can be calibrated out of the measurement \cite{Ranzani2013,Stanley2022}, and since our conclusions are independent of the inclusion of the microwave background, we set $a$ = 1, $\alpha = 0$, $\tau = 0$ in the first section for analytical simplicity. In later sections, when real measurements are modeled for comparison, these microwave background parameter values will be reconsidered.

To better represent actual resonator measurements, random noise is injected into the artificial resonance data to simulate the effects of measurement noise. Two different types of noise are included. The first and simplest type of noise is Gaussian noise from thermal fluctuations \cite{Johnson1928,Nyquist1928} in the resistive components of instruments and measurement setup. We assume that this noise is complex and scatters not only the amplitude of the signal but also its phase by some small random degree. In analogy to a more general mathematical study \cite{Al-Sharadqah2009} on circle fits, the complex thermal noise is injected into the artificial signal for each frequency point $f_i$ as $S_{21}(f_i) = \tilde{S}_{21}(f_i) + \delta_i + i\epsilon_i$, where $\delta_i$ and $\epsilon_i$ are each random variables selected from a Gaussian distribution with mean zero and standard deviation $\sigma_n$.

In addition to thermal noise, the resonance itself exhibits frequency (or phase) noise due to the presence of TLS fluctuators \cite{Gao2007, Barends2008}. Such noise follows a $1/\sqrt{f}$ spectral distribution and can be modeled by small fluctuations of the resonance frequency $f_r$ which are caused by time-dependent corrections to the effective dielectric constant of the substrate. In this work, the strength of this effect will be parameterized by the standard deviation of the resonance frequency $\sigma_{f_r}$, which for the resonance measurements shown here is on the order of tens of Hz. On the complex plane, this noise manifests as a slight rotational jitter for data points on the circle's circumference. As discussed in Appendix C, the inclusion of such frequency noise is necessary for accurately modeling measurements of superconducting resonators.

One of the goals of this study is to determine the reliability of resonance quality factor measurements using a diameter correcting circle fit in the extreme limits where $Q_i\ll Q_c$ and $Q_c\ll Q_i$. To do this, artificial resonance signals can be generated and then fit to determine fit error. The resonance signals are generated with known or "true" internal quality factors $\tilde{Q}_i$, which can be compared to the $Q_i$ returned from the circle fit process. The algorithm from Ref.~\onlinecite{Probst2015} is used to determine $Q_i$ of the artificial $S_{21}$ data. A series of artificial signals is generated with varying $Q_i$ factors with a fixed $Q_c = 10^4 e^{-i\phi}$ with $\phi = \pi/6$ corresponding to a small impedance mismatch \cite{Khalil2012, Probst2015} between the resonator and the measurement line. Because the linewidth $\Delta f_r = f_r/Q_l$ of the resonance signal changes several orders of magnitude, the frequency span $\Delta F$ for the simulated data must also be adjusted to accommodate very high quality resonance signals. As we will show in section \ref{biassection}, changes in the ratio $\Delta F/\Delta f_r$ can affect the fitting results. For each resonance signal, a frequency span encompassing ten times the resonance linewidth was chosen (i.e. $f \in [f_r - 5\Delta f_r, f_r + 5\Delta f_r]$) to ensure an appropriate fitting range for the resonance signal.

\begin{figure*}
\includegraphics[scale = 0.55]{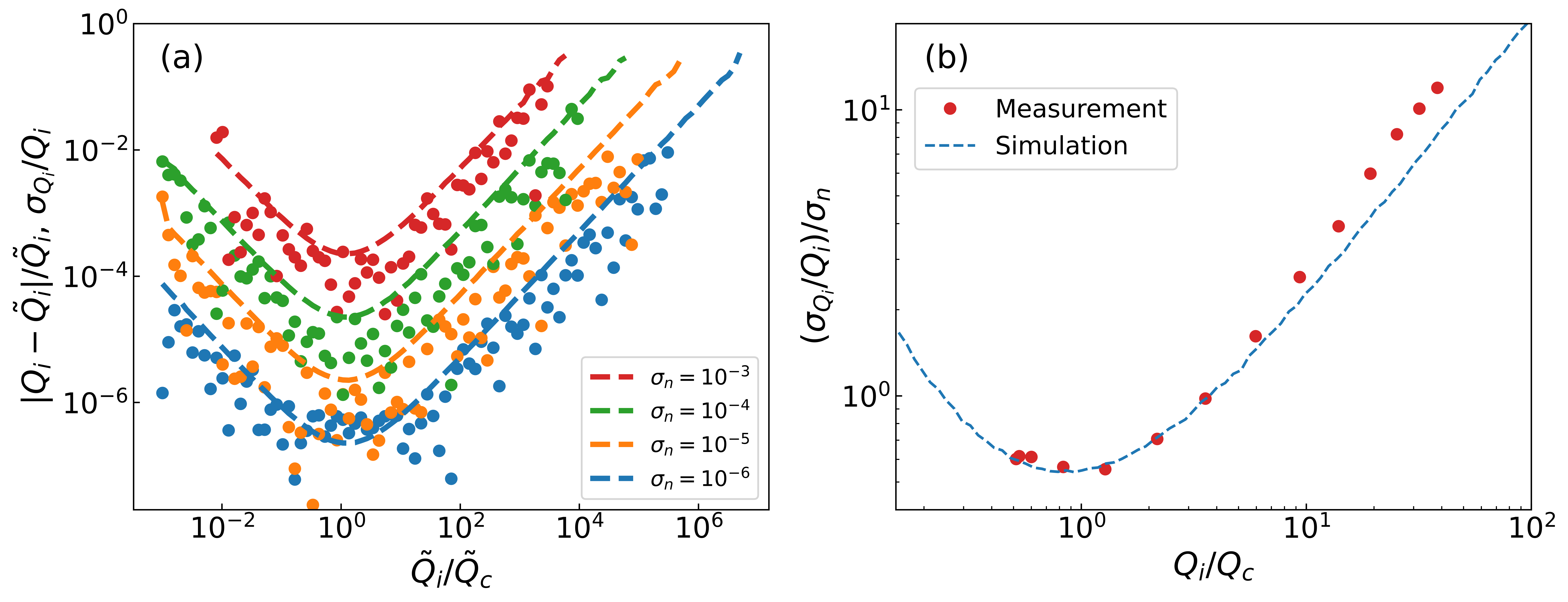}
\caption{Fitting errors on the internal quality factor $Q_i$ for simulated resonance data. Simulations assume an ideal or calibrated microwave background ($a = 1$, $\alpha = \phi = 0$, $\tau = 0$), coupling quality factor $\tilde{Q}_c = 10^4$, and resonance frequency $f_r = 5$~GHz. $\tilde{Q}_i$ is varied from $10$ to $10^{11}$ and $N = 20001$. The measurement bandwidth is maintained at $\Delta F = 10\Delta f_r$, where $\Delta f_r = f_r/Q_l$, and is centered around $f_r$. Only complex Gaussian noise with standard deviation $\sigma_n$ is injected into the artificial signal ($\sigma_{f_r} = 0$). Such noise corresponds to noise in the transmission measurement background. (a) The effects of this noise on the relative fitting error $\sigma_{Q_i}/Q_i$ (dashed lines) and the true relative error $|Q_i - \tilde{Q}_i|/\tilde{Q}_i$ (solid points, corresponding colors). There is close agreement between fit and actual errors, indicating that fit error calculations adequately describe the true error. (b) $\sigma_{Q_i}/Q_i$ for the data in Fig.~\ref{fig:powerdep} (red dots). Because measurement noise is power dependent, the fit error is normalized by the background noise level $\sigma_n$, which is estimated as the standard deviation of data farthest from $f_r$. The dashed blue line represents a fit of artificial resonance data with identical fit and measurement parameters. \label{fig:qi_err} }
\end{figure*}

Figure~\ref{fig:qi_err}(a) shows a comparison of fit $Q_i$ values compared to the true $\tilde{Q}_i$ values used to generate the artificial signals for a range of $\tilde{Q}_i/\tilde{Q}_c$. The relative error $|Q_i-\tilde{Q}_i|/\tilde{Q}_i$ is the deviation of the fit-determined $Q_i$ values from the true $\tilde{Q}_i$ and can be measured across a range of coupling ratios $\tilde{Q}_i/\tilde{Q}_c$. The behavior of the error is non-monotonic with a minimum near critical coupling $\tilde{Q}_i/\tilde{Q}_c=1$. Away from critical coupling, the error increases linearly with $\tilde{Q}_i/\tilde{Q}_c$. Figure~\ref{fig:qi_err}(a) also shows the calculated relative fit error $\sigma_{Q_i}/Q_i$. As shown, the error curves closely follow the behavior of the relative error, and for most cases track with the upper limit of the relative error. This indicates that the calculated errors of the circle fit adequately cover the range where $\tilde{Q}_i$ can be found. We note that in Fig.~\ref{fig:qi_err}(a), resonance noise is omitted ($\sigma_{f_r}=0$). As discussed in detail in Appendix C, when resonance noise is comparable to the background noise, fit errors are increased and overestimate $|Q_i-\tilde{Q}_i|/\tilde{Q}_i$. Therefore, while resonance noise increases fit errors, such an increase does not reflect the true error of the fit. Such an overestimation of the fitting error should be universal for all fitting methods that do not distinguish and separate phase and radial degrees of freedom when calculating fit uncertainties.

The relative error can also be studied as a function of the noise level $\sigma_n$. As shown in Fig.~\ref{fig:qi_err}, the fit errors increase with increasing noise levels. Indeed, we find that the error on $Q_i$ scales geometrically with the measurement noise $\sigma_n$ (see Appendix C). Indeed, when the error is normalized by $\sigma_n$, they fall onto a common curve (Fig.~\ref{fig:qi_err_S}(a)). The same behavior is found when the errors are normalized by $N^{-1/2}$, where $N$ is the number of data points used for the fitting process. Therefore, the error is found to behave as $\sigma_{Q_i}(Q_i/Q_c,\sigma_n,N) = (\sigma_n/\sqrt{N}) \sigma_{Q_i}^*(Q_i/Q_c)$, where $\sigma_{Q_i}^*(Q_i/Q_c)$ is a function independent of both the noise level $\sigma_n$ and number of data points $N$. This result is consistent with a more general analysis for algebraic circle fit errors \cite{Al-Sharadqah2009}. The geometric dependence of the uncertainty on $\sigma_n/\sqrt{N}$ implies that, independent from the coupling regime, the uncertainty on a fit of $Q_i$ can be reduced by either decreasing the noise level (e.g. through trace averaging or filtering) or by increasing the number of data points of the measurement.

\section{Measurements}
The results shown in the Fig.~\ref{fig:qi_err}(a) have been verified experimentally using a measurement of a CPW resonator fabricated from an NbN film on a Si substrate. We take advantage of the power and temperature dependence of the internal quality factor for a superconducting resonator. At the lowest temperatures, resonance losses are dominated by two-level system (TLS) losses at the dielectric interface \cite{McRae2020, Wolz2021}. Such interface TLSs absorb photons from the resonator, creating a loss mechanism and increasing $\Delta f_r$. At low input power, where the average photon number is low, TLS losses are maximal, but as power increases, TLS states become persistently excited by the greater number of resonator photons. As a result, TLS losses decrease, and the resonator's internal quality factor increases. This is indeed observed experimentally, as shown in the inset of Fig.~\ref{fig:powerdep}, which shows the internal quality factor of an NbN resonator.

Using the power dependence of the resonator internal quality factor, the coupling ratio $Q_i/Q_c$ can be tuned. Since the coupling quality factor is affected only by the resonator design and the resonance harmonic \cite{Gopple2008}, it has no expected power dependence, as is consistent with the results of the fitting algorithm. By sweeping input power to the resonator, $Q_i$ increases from $Q_i<Q_c$ to $Q_c<Q_i$. As previously noted, the errors on $Q_i$ are affected by the measurement noise $\sigma_n$, which changes with input power $P_{in}$. Trace average number is increased with decreasing power with 920 trace averages at the lowest power (see Appendix B for more details). Since $\sigma_n$ changes from both power and trace averaging, the fit errors are normalized by $\sigma_n$, and the results are shown in Fig.~\ref{fig:qi_err}(b). For comparison, we simulated an artificial resonance signal using the fit parameters of the measured resonator (background, $\phi$, $Q_c$, etc.) while tuning $\tilde{Q}_i$. We find good agreement between the analyses of our artificial and real resonance data, demonstrating that the simulated data adequately imitates real measurement data.

Since many of the results here focus on the circle fit errors, we detail here how such uncertainty values are calculated. Additional information can be found in Ref.~\onlinecite{Probst2015} and the qkit code \cite{qkit}. Circle fit errors themselves are calculated as $\chi_i = S_{21}^{data}(f_i)-S_{21}^{fit}(f_i)$. A Jacobian matrix $\hat{J}$ is calculated with components $J_{ij}=Re\{\frac{dS_{21}}{d\epsilon_j}\frac{\chi_i}{|\chi_i|}\}$ with $\epsilon_j$ respectively being $Q_l$, $|Q_c|$, $f_r$, and $\phi$ in the four columns of the matrix. With the total squared error $\chi^2 = \sum|\chi_i|^2$, the covariance matrix is given as $\hat{\sigma}^2 = \frac{\chi^2}{N-4}(\hat{J}^{T}\hat{J})^{-1}$, where $\hat{J}^{T}$ is the transpose of $\hat{J}$. The uncertainties on $Q_l$, $|Q_c|$, $f_r$, and $\phi$ are then taken as the square roots of the diagonal components of $\hat{\sigma}^2$. The error on $Q_i$ is then derived from the uncertainties on $Q_l$ and $Q_c$ using standard error propagation techniques.

\section{Fitting Bias}
\label{biassection}

\begin{figure*}
\includegraphics[scale = 0.5]{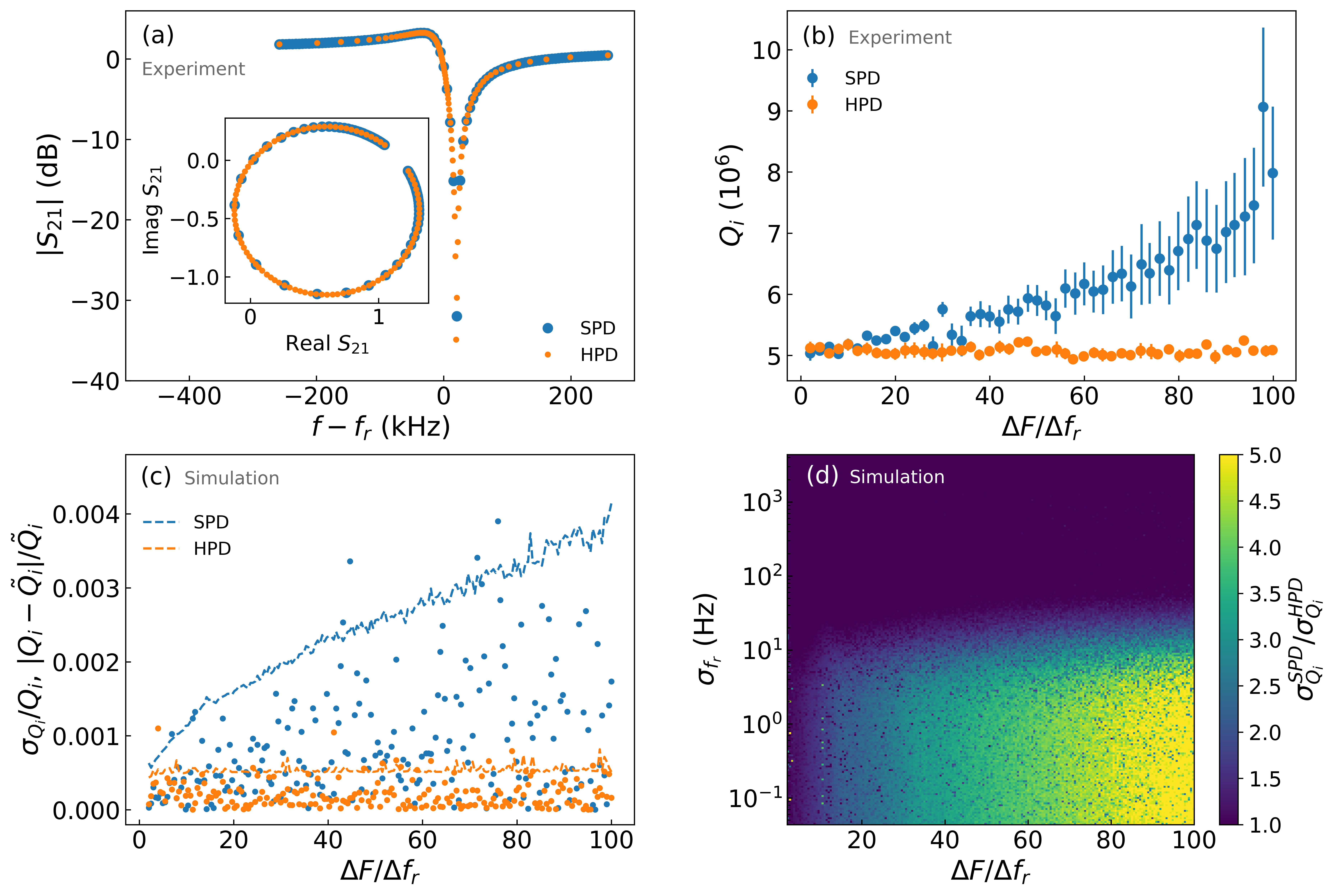}
\caption{(a) A comparison of two measurements of the same resonator signal using the SPD and HPD ($N=101$, $\Delta F/\Delta f_r = 10$). (inset) The two data sets project onto the complex plane as a circle. Whereas the SPD has an increased point density near the off-resonant point, the HPD has a uniform point density around the circumference of the circle. (b) Fit values for $Q_i$ using the SPD and HPD measurements shown in (a) except with $N=10001$. The measurement bandwidth $\Delta F$ was gradually increased as a means of varying the degree of bias. Aside from the measurement band and point distributions, all other measurement parameters were identical. Whereas the SPD shows a gradual bias toward higher $Q_i$ values, the fit of HPD returns a constant $Q_i$, independent of $\Delta F$. (c) As in Fig.~\ref{fig:qi_err}(a), a systematic fit of simulated data allows for a comparison of the fit uncertainty (dashed lines) and the true error (points). While the SPD error increases as $\sqrt{\Delta F}$ (see Appendix D), the HPD error remains constant at the minimal error of the SPD. In other words, the HPD maximizes fit accuracy relative to the SPD. (d) The increased performance of the HPD over the SPD is dependent on the level of resonance frequency noise $\sigma_{f_r}$. The relative performance can be parametrized by the error ratio $\sigma_{Q_i}^{SPD}/\sigma_{Q_i}^{HPD}$, where $\sigma_{Q_i}^{SPD}$ and $\sigma_{Q_i}^{HPD}$ are the errors from the SPD and HPD fits, respectively. At low $\sigma_{f_r}$, $\sigma_{Q_i}^{SPD}/\sigma_{Q_i}^{HPD}$ follows the behavior shown in (c) for the case with $\sigma_{f_r}=0$. However, near $\sigma_{f_r}>10$~Hz, the frequency noise begins to dominate the fit errors and $\sigma_{Q_i}^{SPD}/\sigma_{Q_i}^{HPD}$ rapidly approaches 1. This behavior implies that the benefit of the HPD over the SPD is due to a reduction of $\sigma_n$ error contributions.\label{fig:hpd}}
\end{figure*}

In the standard vector network analyzer (VNA) transmission measurement, frequency data points are distributed linearly across the measurement span $\Delta F$. However, when a resonance measurement with this standard point distribution (SPD) is projected onto the complex plane, points are distributed inhomogeneously over the circumference of the circle with a qaudratically increasing point density near the off-resonant point (see Appendix D for a derivation). As has been shown for the fitting of power-laws \cite{Newman2005}, nonuniform point distributions can potentially lead to bias in fitting algorithms. To further understand if such biases exist in the circle fit of SPD measurements, we have constructed an alternative homophasal point distribution (HPD), which redistributes the frequency measurement points such that the resonance data is homogeneously distributed around the complex circle (i.e. homogeneous in phase). This redistribution method is similar to one derived for nanomechanical resonators \cite{SarsbyThesis}. An example comparison of SPD and HPD resonator data is shown in Fig.~\ref{fig:hpd}(a).

The HPD is constructed using the frequency-phase relation \cite{Probst2015} $\theta(f) = \theta_0 + 2\arctan(2Q_l[1-f/f_r])$, which depends only on the loaded quality factor $Q_l$, the resonance frequency $f_r$, and an offset phase $\theta_0$ (see Ref.~\onlinecite{Probst2015} for further details). The inverse function is $f(\theta) = f_r(1-\frac{1}{2Q_l}\tan[\frac{\theta-\theta_0}{2}])$, and by defining $N$ equidistant points distributed within $\theta-\theta_0 \in (-\pi,\pi)$, a new frequency point distribution can be calculated that will spread points uniformly around the circle's circumference. Importantly, this method does not require an explicit prior determination of $Q_c$ or $Q_i$ and only an approximate determination of $Q_l$, $f_r$, and $\theta_0$. It should be noted that a least-squares fit of the phase data is already used in the circle-fitting process of Ref.~\onlinecite{Probst2015} to determine these parameters.

Fitting errors are analyzed in the same manner as the $Q_i/Q_c$ dependence. An artificial resonance signal was generated with a known $\tilde{Q}_i$ and $\tilde{Q}_c$, while the frequency distribution was generated using both SPD and HPD. Equivalent levels of noise were generated independently and then injected into both signals. To test for bias, the point distribution of the SPD is tuned by changing the measurement span $\Delta F$. As discussed in Appendix D, the point density of the SPD becomes more clustered toward the off-resonant point as $\Delta F$ grows larger. To test for fit bias, the same resonance signal is repeatedly measured with a steadily increasing $\Delta F$ centered around $f_r$. As is shown in Fig.~\ref{fig:hpd}(b), the fit value of $Q_i$ steadily increases with $\Delta F$. Such an increase does not reflect an actual change in the resonator quality and derives purely from the measurement protocol and fit algorithm. In contrast, an HPD measurement remains consistent over a wide range from $1\leq \Delta F/\Delta f_r \leq 100$.

Not only does the HPD remove bias from the measurement, it also minimizes fit error with respect to the SPD. Figure \ref{fig:hpd}(c) shows the $Q_i$ fit errors for the SPD and HPD on an artificial overcoupled resonator signal. While the SPD error increases as $\sigma_{Q_i}^{SPD} \propto \sqrt{\Delta F/\Delta f_r}$, the error on the HPD remains constant, demonstrating that HPD measurements are independent from the choice of measurement span. The constant $\sigma_{Q_i}^{HPD}$ value also corresponds to the minimum $\sigma_{Q_i}^{SPD}$ at $\Delta F/\Delta f_r = 1$, and therefore the HPD is a more efficient measurement protocol than the SPD. This benefit of the HPD over the SPD is not independent of the resonator properties, however. Figure \ref{fig:hpd}(d) shows a colormap of the error ratio $\sigma_{Q_i}^{SPD}/\sigma_{Q_i}^{HPD}$ as a function $\Delta F/\Delta f_r$ and resonance frequency noise level $\sigma_{f_r}$. When $\sigma_{f_r}$ is small, the errors of the SPD and HPD behave as described in Fig.~\ref{fig:hpd}(c), but as resonance noise becomes the dominant source of error, $\sigma_{Q_i}^{SPD}/\sigma_{Q_i}^{HPD} \approx 1$, and therefore in this regime there is no benefit of the HPD over the SPD. However, while there are indeed regimes where errors of the HPD match those of the SPD, it should be emphasized that in no regime is the SPD observed to outperform the HPD.


\section{Discussion}
\begin{figure}
\includegraphics[scale = 0.45]{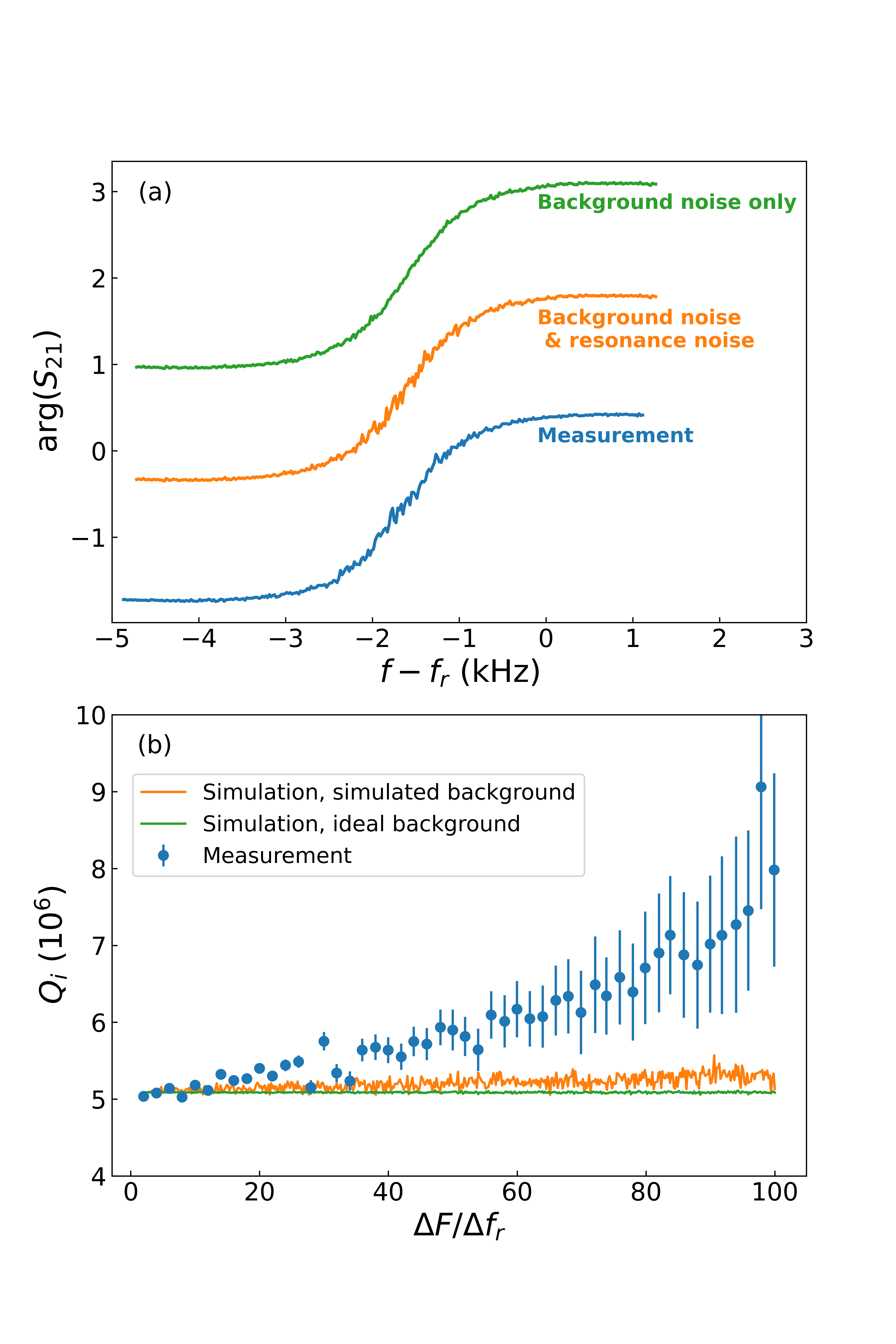}
\caption{(a) A comparison of artificial and measurement $\arg (S_{21})$ data as a function of frequency. The top curve (green) is artificial data with only complex background noise $S_{21} = \tilde{S}_{21} + \delta + i\epsilon$ added. The middle curve (orange) is generated with both background noise and frequency noise $\sigma_{f_r}\approx 50$~Hz. The bottom curve (blue) is the phase data for the -35~dBm measurement of Fig.~\ref{fig:powerdep}. The two artificial resonance curves are generated with fit parameters derived from a circle fit of the measurement curve. The curves are displaced vertically for clarity. As shown here, the inclusion of $f_r$ noise is essential for adequately replicating measurement data. (b) A comparison of fit results for simulated and measured resonance data. Blue data points are $Q_i$ fit values for the resonator shown in Fig.~\ref{fig:hpd}(a) as a function of span ratio $\Delta F/\Delta f_r$. The orange curve is a fit result on simulated data with parameters identical to the measurement data with $|Q_c| \approx 6.730 \times 10^4$, $Q_i \approx 5.181\times 10^6$, $\phi \approx 0.668$, $f_r \approx 4.364$~GHz, $a \approx 1.149$, $\alpha \approx 1.597$, and $\tau \approx -8.825\times 10^{-11}$~s. The noise levels are $\sigma_n = 0.00035 + i 0.00016$ and $\sigma_{f_r} = 50$~Hz and comparable to estimates of the noise in the measurement data. The green curve is a similar simulation, except with an ideal background with $a = 1$, $\alpha = 0$, and $\tau = 0$. \label{fig:freqnoise}}
\end{figure}

As demonstrated above, we find that the circle fitting procedure of Probst et al. \cite{Probst2015} accurately determines $Q_i$ over a wide parameter space. While this study used superconducting notch-type resonators, the methods used here are general and can be applied to any type of resonance system. Although fit errors increase linearly with $|Q_i/Q_c|$, the fit values are observed to remain within a standard deviation of the true values over the full range of coupling parameters. Depending on the noise source, fit errors either accurately predict the true error of the fit or overestimate it slightly. Although $Q_i$ fit values are still found within the $95\%$ confidence interval, frequency noise increases the calculated fitting error without a corresponding increase in true error (see discussion in Appendix C). The frequency noise leads to rotational error that increases $\chi^2$ but does not affect the calculation of $Q_i$, which is determined only by the radial degrees of freedom. Interestingly, like the systematic errors induced by parasitic Fano resonances in the measurement circuit \cite{Rieger2023}, the effect from frequency noise is strongest in the overcoupled regime (see Fig.~\ref{fig:freqnoise}d), implying that this regime is particularly susceptible to increased errors. Regardless of the regime, this uncertainty overestimation from frequency noise should be present for all fit algorithms that do not distinguish between phase and radial degrees of freedom when calculating fit uncertainties. Therefore, a fit uncertainty calculation that isolates only the radial degrees of freedom may lead to lower errors.

On the other hand, we find that a new type of measurement protocol, specifically one possessing a homophasal point distribution, can reduce fitting errors by redistributing points around the resonance circle. Errors for standard VNA measurements (i.e. the SPD) grow with the square-root of the ratio $\Delta F/\Delta f_r$. Therefore, for a typical measurement with $\Delta F \gtrsim 4\Delta f_r$, a factor of two or greater improvement in accuracy can be expected for circle fit results. Interestingly, SPD errors are minimized when $\Delta F/\Delta f_r = 1$ (errors grow dramatically for $\Delta F/\Delta f_r < 1$ for both the SPD and HPD). However, when $\Delta F = \Delta f_r$, this is only a measurement of half of the resonance circle \cite{Megrant2012}, implying that under some circumstances a half-circle fit can return better accuracy than a full circle fit for a constant $N$. An improvement on the fitting error, either by implementing an HPD or by changing $\Delta F/\Delta f_r$, allows for the possibility for reducing measurement time without a sacrifice accuracy. By implementing these methods, the total number of data points, and therefore time per measurement, can be reduced while maintaining a constant the same accuracy. Indeed, since errors scale inversely with $\sqrt{N}$, a factor of two error reduction from an HPD would allow for a factor of four reduction in $N$ and time to maintain an equivalent error with an SPD measurement.

Insight into why the HPD outperforms the SPD can be gained by considering the information that is contained within a microwave transmission measurement. Appendix F details how the information yield of a resonator measurement can be defined in terms of the Shannon entropy $H(f)$ \cite{Shannon1948}. As shown in Fig.~\ref{fig:entropy}(a), the $H(f)$ peaks at $\pm\Delta f_r/2$ for an ideal resonance signal, meaning that these points nearest to the resonance linewidth are the most uncertain and therefore contain the most information about the resonance signal once measured. Indeed, as shown in Fig.~\ref{fig:entropy}(b), the total entropy $H_{set}$ within a measurement sweep peaks near $\Delta F/\Delta f_r \sim 1.5$. The HPD maintains an overall higher information content compared to the SPD, and this corresponds to the decreased fit error (inset Fig.~\ref{fig:entropy}(b)). 

We note that this simplistic analysis of the information yield only considers amplitude degrees of freedom. Furthermore, the mutual information \cite{Kraskov2004} shared between measurement points may need to be considered when the density of points becomes very high, such as in the regime $\Delta F < \Delta f_r$, and correlations between each adjacent measurement point becomes large. A formal theory of the information from a resonance signal measurement may need to include additional degrees of freedom (such as phase/frequency) and the influence of the information shared between data points. The establishment of such a formal theory will be the goal of future work.

As demonstrated, the circle fit is also susceptible to bias from changes to the measurement span $\Delta F$. At relatively modest values of $\Delta F/\Delta f_r \gtrsim 10$, a systematic trend which exceeds the error can be observed in the data of Fig.~\ref{fig:hpd}(b). For artificial resonance data with an ideal background, we find no bias with $\Delta F$ (see Appendix C), indicating that the true source of bias is not the point distribution of the SPD but rather how this distribution weights in favor of influences from the microwave background signal. Because the HPD shifts data points to focus on the resonance signal itself, it minimizes the influence of the microwave background on the fitting procedure and thus the bias. Furthermore, while a small degree of bias can be replicated in our simulated data, the degree of bias is much smaller than found in the actual measurement. This indicates that the majority of bias originates from sources not captured in the model, such as potential parasitic resonances in the measurement setup \cite{Rieger2023}, and while some works \cite{Noguchi2019, Zollitsch2019} have indeed presented $Q$-factor results for measurements with $\Delta F/\Delta f_r > 10$, the more important and broader conclusion from the study presented here is that since the degree of fit bias is unknown, a systematic measurement protocol needs to be implemented to ensure such biases do not occur even for the case of $\Delta F/\Delta f_r < 10$.

The point distribution of the SPD has a more complicated dependence on $\Delta f_r$ (see Appendix D). In Fig.~\ref{fig:qi_err}(b), a slight deviation of the measurement and simulated data is observed. We postulate that this deviation is due to the $\Delta f_r$ dependence of the SPD point distribution. Since the point distribution of SPD measurements changes with the resonance linewidth, bias may influence results for wideband measurements with changing resonance quality factors \cite{Song2009, Diener2012, Coumou2013a, Carter2019, Xu2019}. We want to emphasize that none of the studies cited here should be regarded as affected by bias, since bias in this study is only demonstrated for one specific measurement setup and fitting algorithm. It is unknown if other fitting algorithms (e.g. those presented in Refs.~\cite{Khalil2012, Megrant2012, Carter2017, Shao2013}) are susceptible to the same sources of bias, however this possibility may explain the high variability of results observed in a comparison of fitting methods \cite{Mcrae2022}. Because of this, care should be taken when comparing $Q$-factor measurements across studies with different measurement and fitting procedures for the sake of determining an optimal materials platform for quantum computing.


\section{Conclusions}
As demonstrated here, circle fit analysis remains a robust method for characterizing resonance signals. Even in extreme limits where $Q_i \gg Q_c$, the method can return an accurate measurement of $Q_i$. Away from critical coupling, where $\sigma_{Q_i}$ reaches a minimum, the errors on $\sigma_{Q_i}$ increase only linearly with $Q_i/Q_c$. Errors can be mitigated by increasing the number of measurement points and reducing signal noise but have a lower limit due to the intrinsic resonance noise from TLS effects. Furthermore, we have related the fit uncertainty to the information content within a data set and have shown that measurements that are optimized to have more information produce more accurate predictions on $Q_i$.

On the other hand, bias in circle fit results due to the influence of the microwave background has also been demonstrated. By comparing simulation to actual measurements, we find that the degree of bias could not be replicated in the standard resonator model. Therefore, we conclude that a systematic study is required to ensure that such biases are not affecting results from other fitting methods. To resolve this issue for the circle fit method, we have established a new measurement technique (the HPD) that minimizes not only fit bias but also the overall fitting error. For a typical resonance measurement with $\Delta F \sim 4\Delta f_r$, we expect a factor of two fit error reduction. Therefore, we conclude that the HPD is an overall improved method for resonance measurement compared to the standard method. Due to its generality, this new protocol can be implemented not only for the study of superconducting notch-type resonators but also other resonance systems such as those in the fields of spin-waves and optics.

\begin{acknowledgments}

This work was supported by Innovate UK (QTools, grant No 79373 and FABU, grant No 50868); the EPSRC Oxford Quantum Computing Hub EP/T001062/1; and the FET Open inititative from  the European Union's Horizon 2020 program under N$^\circ$ 899561. We would also like to thank Russ Renzas and Harm Knoops at Oxford Instruments Plasma Technology for advising on the manuscript and providing fruitful discussions on resonator and thin film characterization.

\end{acknowledgments}



\begin{figure*}
\includegraphics[scale = 0.5]{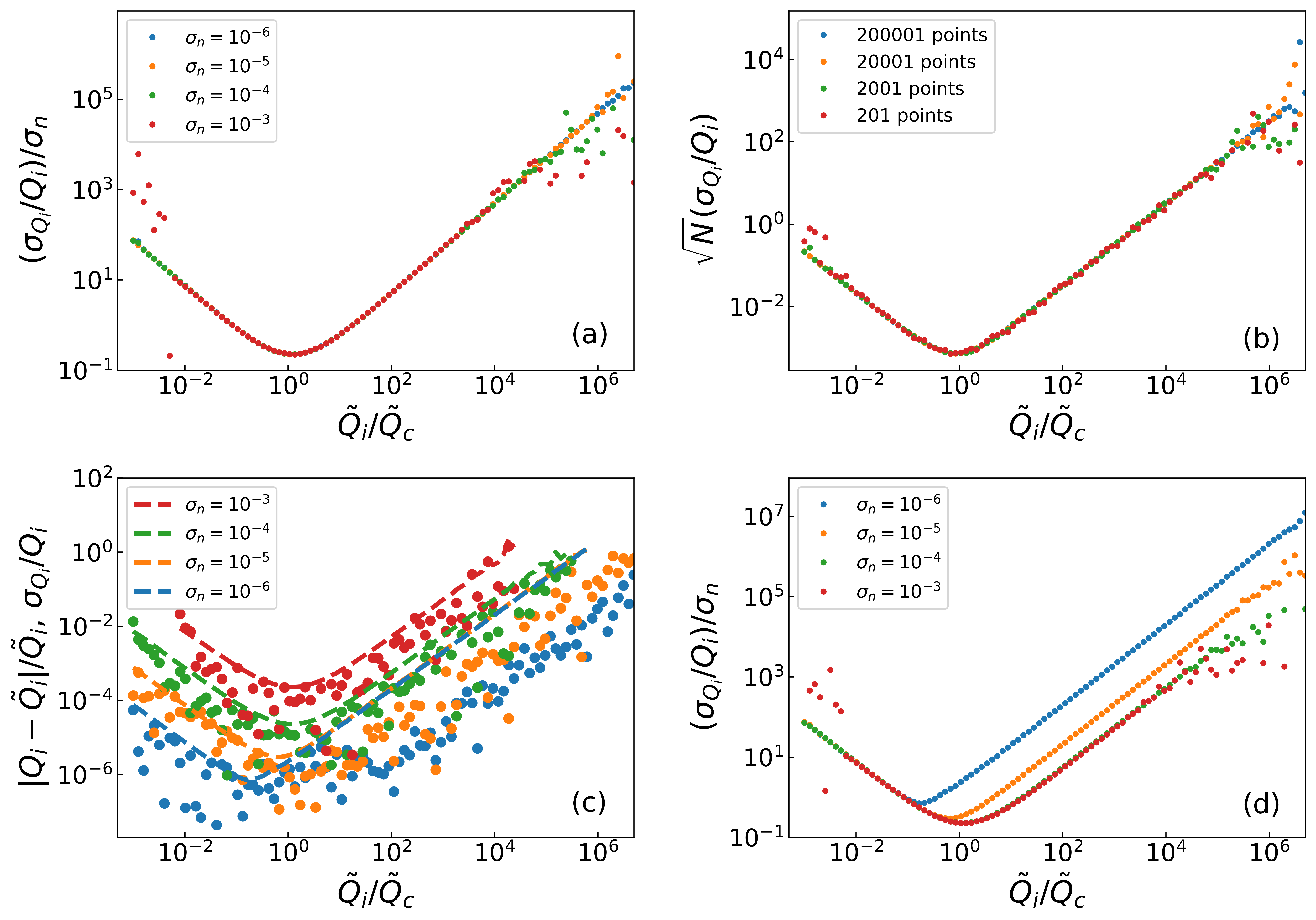}
\caption{The fitting errors on the internal quality factor $Q_i$ for simulated resonance data. Simulations assume an ideal or calibrated microwave background ($a = 1$, $\alpha = \phi = 0$, $\tau = 0$), coupling quality factor $\tilde{Q}_c = 10^4$, and resonance frequency $f_r = 5$~GHz. $\tilde{Q}_i$ is varied from $10$ to $10^{11}$. The measurement bandwidth is maintained at $\Delta F = 10\Delta f_r$, where $\Delta f_r = f_r/Q_l$, and is centered around $f_r$. Except where noted otherwise, $N = 20001$. (a) Shows fitting errors for various background noise levels $\sigma_n$, similar to the dashed curves of Fig.~\ref{fig:qi_err}(a). Frequency noise is omitted ($\sigma_{f_r}=0$~Hz). When errors are normalized by $\sigma_n$, they collapse onto a single curve as shown. This indicates that errors scale linearly with the background signal noise. (b) $Q_i$ errors for fits with different number of points $N$. When multiplied by $\sqrt{N}$, errors also collapse onto a single curve. Therefore, as noted in the main text, $\sigma_{Q_i}/Q_i \propto \sigma_n/\sqrt{N}$. (c) Resonator noise can also be added into the simulated data. This is done by adding a small random deviation ($\sigma_{f_r}\approx 50$~Hz) to the resonance frequency for each data point calculated with Eq.~\ref{resEq}. The frequency noise increases the fit error (dashed lines) in the overcoupled regime, however, the frequency noise does not affect the true error (scatter points). (d) When normalized by $\sigma_n$, the errors in (c) scale with $\sigma_n$ for large values of background noise, but this scaling breaks down for small $\sigma_n$.  \label{fig:qi_err_S}}
\end{figure*}

\section{Appendix A: Table of Variables}

To assist the reader, we have included a table  of variables and definitions used to model the resonance signal and noise.

\begin{table*}[t]
\centering
\resizebox{\textwidth}{!}{
\begin{tabular}{|l|l|}
\hline
\hfil Variable & \hfil Definition\\
\hline
 \hfil $\tilde{S}_{21}$, $S_{21}$ & $\tilde{S}_{21}$ is an ideal complex transmission resonance signal with zero measurement noise but potentially affected by the measurement circuit. $S_{21}$ is a resonance signal under the influence of measurement noise.\\
\hline
 \hfil $\tilde{Q}_i$, $Q_i$ & The internal quality factor. $\tilde{Q}_i$ is used to calculate $\tilde{S}_{21}$ in the generation of artificial resonance data. $Q_i$ is the measured/fitted value.\\
\hline
 \hfil $\tilde{Q}_c$, $Q_c$ & The complex coupling quality factor $Q_c = |Q_c|e^{-i\phi}$. $\tilde{Q}_c$ is used to calculate $\tilde{S}_{21}$ in the generation of artificial resonance data. $Q_c$ is the measured/fitted value. \\ 
\hline
 \hfil $\sigma_{Q_i}$ & The fit error on $Q_i$. $\sigma_{Q_i}^{SPD}$ and  $\sigma_{Q_i}^{HPD}$ denote values specified for SPD and HPD distributions.\\
\hline
 \hfil $\phi$ & The phase shift of the resonance signal due to impedance mismatch.\\
\hline
 \hfil $\tilde{Q}_l$ & The loaded quality factor $Q_l^{-1} = Q_i^{-1} + \mathrm{Re}\{Q_c^{-1}\}$.\\
\hline
 \hfil $f$, $f_i$ & The signal frequency used to probe the transmission response. For a set of measurements, $f_i$ is the $i$th frequency of the set.\\
\hline
 \hfil $\theta$ & The complex phase of $S_{21}$.\\
\hline
 \hfil $N$ & The number of data points in the measurement set.\\
\hline
 \hfil $\Delta F$ & The frequency span of the set of the frequencies in the measurement set.\\
\hline
 \hfil $f_r$, $\Delta f_r$ & The resonance frequency and the resonance linewidth: $\Delta f_r = f_r/Q_l$. \\
\hline
 \hfil $\sigma_{f_r}$ & The standard deviation of the frequency noise introduced to $f_r$. \\
\hline
 \hfil $a$, $\alpha$, $\tau$ & The background transmission amplitude, phase offset, and microwave delay, all of which arise from the specifics of the measurement setup. \\
\hline
 \hfil $\delta_i$, $\epsilon_i$ & Random variables added to $\tilde{S}_{21}$ calculated at $f_i$ and generated on a Gaussian distribution with a zero mean. $\delta_i$ is the real component of the noise and $\epsilon$ is the imaginary component of the noise.\\
\hline
 \hfil $\sigma_n$ & The standard deviation of the Gaussian distribution used to generate $\delta_i$ and $\epsilon_i$.\\
\hline
\hfil $N$ & The number of data points within a VNA frequency scan.\\
\hline
\hfil $N_{tr}$ & The number of traces performed by a VNA during a measurement.\\
\hline
\hfil $P_{VNA}$, $P_{chip}$ & The microwave power applied by the VNA during the measurement and the power applied to the chip under probe, respectively.\\
\hline
\hfil $H(f)$ & The Shannon entropy of a probabilistic measurement.\\
\hline
\hfil $H_{set}$ & For a VNA scan of a set of $N$ points, $H_{set}$ is the total information entropy of the set.\\
\hline
$\langle n_{r} \rangle$, $\langle n_{ph} \rangle$, $\langle n_{ph}' \rangle$ & The number of photons in the resonator, the VNA output signal, and the VNA return input after traversing the system under probe, respectively.\\
\hline

\end{tabular}}
\caption{Table of the variables used in this manuscript and their definitions.}
\end{table*}

\section{Appendix B: Sample Details and Measurement Protocols}
\label{appB}
The superconducting CPW microwave resonators used in this study were fabricated from a 55~nm NbN film deposited on a Si substrate using an Oxford Instruments FlexAL Atomic Layer Deposition system. The resonators were designed with central conductor width $S = 10$~$\mu$m and gap width $W = 6$~$\mu$m. The details of the material growth techniques will be provided in a different manuscript. The designs were etched into the NbN films by an inductively-coupled SF6 plasma on an Oxford Instruments PlasmaPro 100 ICP180, after either optical or electron-beam lithography and respective resist development. The $5\times 5$~mm$^2$ chips were mounted into a microwave measurement box, which itself was mounted onto the mixing chamber stage of a dilution refrigerator. Magnetic shielding was provided by a cryoperm shield. The input microwave lines had $\sim 70$~dB of attenuation. After passing through the chip, output signals were amplified by a 4~K high-electron-mobility transistor (HEMT) providing $\sim 40$~dB amplification. Signals were further amplified an additional $20$ or $40$~dB room temperature amplifier depending on the setup used.

Each curve in Fig.~\ref{fig:powerdep} is measured with $N = 4001$ points distributed linearly across the measurement bandwidth $\Delta F$ and a VNA output power $P_{VNA}$. A power-dependent trace averaging protocol was used to mitigate measurement noise at low powers. For $P_{VNA} < -50$~dBm, $N_{tr} = (P_{VNA}+50)^2+20$, where $N_{tr}$ is the number of traces averaged, while $N_{tr} = 20$ for $P_{VNA} > -50$~dBm. Before the circle fit, transmission phase was normalized by subtracting the linear delay.


\begin{figure*}
\includegraphics[scale = 0.55]{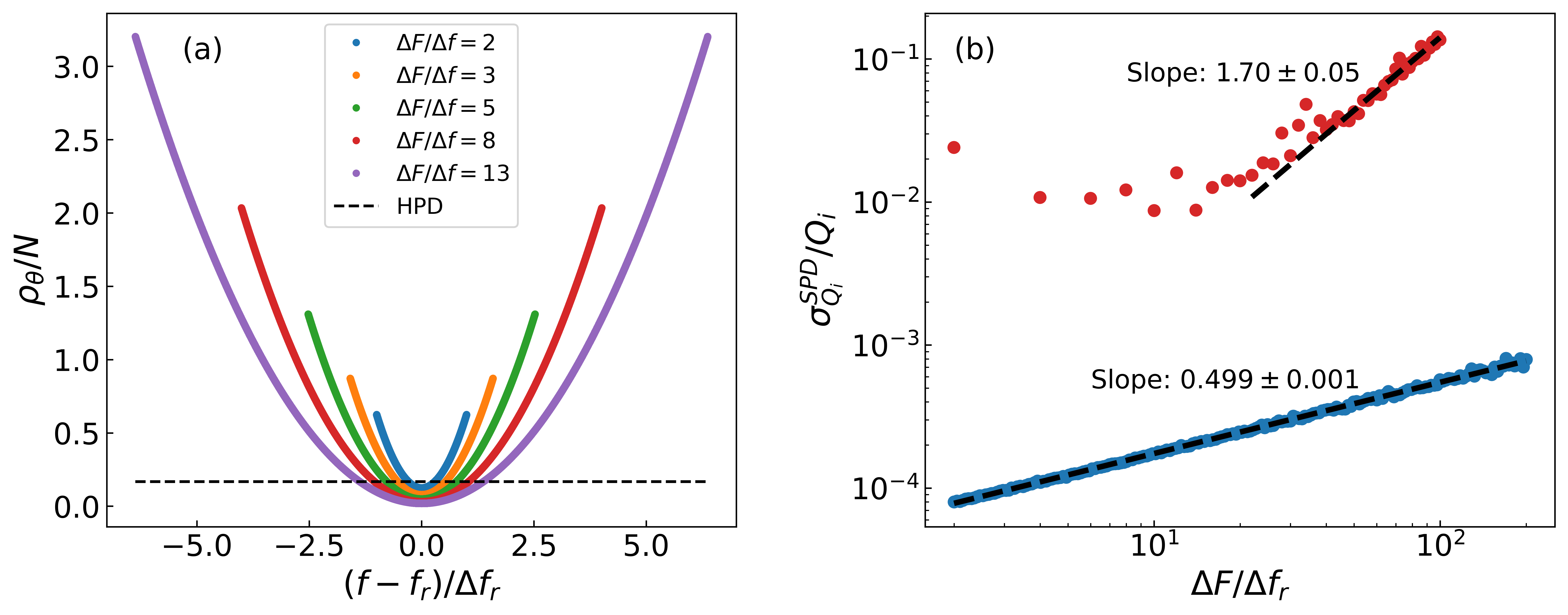}
\caption{ (a) The normalized phasal density of points ($\rho_\theta/N$) around the resonance circle as a function of frequency. These curves are calculated manually from artificial resonance data as $1/\delta \theta$, where $\delta \theta$ is the phase difference between adjacent measurement points. As shown in Eq.~\ref{rhotheta}, the phasal point density of the SPD is quadratic with respect to frequency. As the measurement span $\Delta F$ grows, the density of points near $f_r$ becomes more disperse while data becomes more concentrated toward near the off-resonant point. For comparison, the uniform $\rho_\theta$ of the HPD is shown as a dashed line. (b) A linear fit of the $\sigma_{Q_i}^{SPD}/Q_i$ for artificial resonance data as a function of $\Delta F/\Delta f_r$. A linear fit of the errors show that it grows as $\sigma_{Q_i}^{SPD}/Q_i \propto \sqrt{\Delta F/\Delta f_r}$. This data is similar to those in Fig.~\ref{fig:hpd}(c), except it is generated on a logarithmic scale to prevent fit bias \cite{Newman2005}. \label{fig:phasdens}}
\end{figure*}

\section{Appendix C: Comparison of simulated data with measurements}
\label{appC}
As mentioned in the main text, we included resonance frequency noise to better model real resonator measurements. For more realism, the frequency noise was simulated using a $1/\sqrt{f}$ spectrum, as found in Ref.~\onlinecite{Gao2007}, however no discernible difference in fit results were found when using pink noise rather than white noise. Figure \ref{fig:freqnoise}(a) shows a comparison of artificial resonance ($\arg (S_{21})$) data along with measurement data from Fig.~\ref{fig:powerdep}. As shown, the real measurement data exhibits an increased noise at the phase jump near $f_r$. In simulated data, the standard injection of complex noise does not replicate the increased phase noise shown in the measurement. Instead, by injecting a small random shift on $f_r$ with a standard deviation $\sigma_{f_r}\approx 50$~Hz, an increased phase noise at $f_r$ can be replicated.

The effects of the resonance noise on the circle fit error are shown in Fig.~\ref{fig:qi_err_S}. For $\sigma_{f_r}=0$, fit errors on $Q_i$ agree with the true error $|Q_i-\tilde{Q}_i|/\tilde{Q}_i$ as shown in Fig.~\ref{fig:qi_err}(a). In this case, the errors scale linearly with the background noise $\sigma_n$ as shown in Fig.~\ref{fig:qi_err_S}(a). However, when $\sigma_{f_r}>0$ is included, errors are observed to deviate from this scaling trend. As shown in Fig.~\ref{fig:qi_err_S}(c), $\sigma_{Q_i}/Q_i$ overestimates the true error in the over-coupled regime. Hence, we can conclude that the increase in fit error is only an artificial increase and not a true increase. In addition, errors no longer scale fully with $\sigma_n$. Instead, errors scale with $\sigma_n$ when $\sigma_n$ is large, but this trend deviates for small $\sigma_n$ (Fig.~\ref{fig:qi_err_S}(d)). The source of the increase is likely the increase in $\chi_i$ for each data point due to the rotational jitter from phase noise. However, since the noise is only rotational in nature, it does not actually affect the calculation of the circle diameter and thus $Q_i$. A method that distinguishes radial and phase uncertainties may prove beneficial in reducing fit uncertainty in the presence of large frequency noise.

As discussed in Sec.~\ref{biassection}, fitting results of SPD data have a dependence on the ratio $\Delta F/\Delta f_r$. Interestingly, as shown in Fig.~\ref{fig:freqnoise}(b), the strength of the bias is larger for measured data than for simulated data with identical fit parameters and noise levels. Indeed, for simulated signals with an ideal (i.e. calibrated) background, there is virtually no $\Delta F/\Delta f_r$ dependence for fit results. For simulated data with a non-ideal background, a small but perceptible amount of bias is observed as indicated by a slight slope in the results, however, this bias is still much smaller than the bias (i.e. slope) observed in the measurement data. This indicates both that the types of noise modeled here (frequency and amplitude) do not contribute to fitting bias observed in this study and that the majority of this bias stems from effects not captured in Eq.~\ref{resEq}. The resonator model used here assumes that the background parameters $a$, $\tau$, and $\alpha$ are frequency independent locally near $f_r$, however such assumptions may not be applicable for many cryogenic microwave measurement setups. Furthermore, the influence from Fano resonances \cite{Rieger2023} in the measurement setup is also excluded. Because the source of bias is not captured in the resonator model itself, it is entirely possible for the magnitude of the fit bias to be setup dependent. However, it should be reiterated that the benefit of the HPD is that it removes fit bias even if the source of the bias is seemingly unparameterized and can be implemented without changes to the experimental setup.

\section{Appendix D: Density of points around the circle}
\label{appD}
For the standard point distribution (SPD) of a VNA sweep, data points are distributed quadratically around the circle with a minimum point density near $f_r$ and maximum near the off-resonant point. The distribution can be derived from the frequency-phase relation $\theta(f) = \theta_0 + 2\arctan(2Q_l[1-f/f_r])$ \cite{Probst2015}. By taking the derivative of $\theta(f)$ with respect to $f$ and then inverting (or alternatively by taking the derivative of the inverse function $f(\theta)$ and then substituting $\theta(f)$ back into $df(\theta)/d\theta$), the expression
\begin{equation}
    \frac{df}{d\theta} = \frac{f_r}{4Q_l}\big[4 (1-f/f_r)^2 Q_l^2 + 1\big]
\end{equation}
can be derived. Remembering that $Q_l = f_r/\Delta f_r$, where $\Delta f_r$ is the linewidth, the expression simplifies to \begin{equation}
    \frac{df}{d\theta} = \Delta f_r \bigg[\bigg(\frac{f_r-f}{\Delta f_r}\bigg)^2 + 1/4 \bigg]. \label{eqSabove}
\end{equation}
For the SPD, frequency points are distributed linearly with the separation between points $\delta f = \Delta F/N$, where $\Delta F$ is the frequency span and $N$ is the number of data points. A continuous to discrete transformation can be made assuming a large $N$ limit and by substituting $df/d\theta \to \delta f/ \delta \theta$, where $\delta \theta$ is the discrete difference between neighboring phase data points. With this substitution, the discrete phasal density $\rho_\theta$ of data points around the circle is
\begin{equation}
    \rho_\theta \equiv \frac{1}{\delta \theta} = \frac{N}{\Delta F/\Delta f_r} \bigg[\bigg(\frac{f_r-f}{\Delta f_r}\bigg)^2 + 1/4 \bigg]. \label{rhotheta}
\end{equation}
This expression can be confirmed by manually calculating $1/\delta \theta$ in simulated data, as shown in Fig.~\ref{fig:phasdens}(a). Interestingly, while the distribution is tuned by the measurement span $\Delta F$, it also has a more complicated dependence on the resonance linewidth $\Delta f_r$. Therefore, a constant $\Delta F$ SPD measurement for a resonance signal with a changing linewidth might also exhibit an unknown bias. However, if such biases exist, the HPD protocol can be expected to eliminate such effects by normalizing the point density around the circle. 

Aside from bias, SPD fit errors also increase with increasing $\Delta F/\Delta f_r$, as shown in Fig.~\ref{fig:hpd}(c). Fig.~\ref{fig:phasdens}(b) shows similar data generated on a logarithmic scale. A linear fit of the errors demonstrates how the error $\sigma_{Q_i}^{SPD}$ increases with the square root of $\Delta F/\Delta f_r$. This simulation result conflicts with the experimental results, which find that errors increase as $\sigma_{Q_i}^{SPD} \propto (\Delta F/\Delta f_r)^{1.7}$. This difference between simulation and experimental results further emphasizes how the model in Eq.~\ref{resEq} may not fully capture all sources of error in real experiments. Other sources of errors, such as the inclusion of parasitic Fano resonances in the measurement circuit \cite{Rieger2023} may be necessary to appropriately model resonance signals.


\section{Appendix E: Calculation of Photon Numbers for $\phi>0$}
\label{appE}
In the inset of Fig.~\ref{fig:powerdep}, the internal quality factor $Q_i$ is plotted against the estimated average photon number $\langle n_r \rangle$ in the resonator. The calculation of $\langle n_r \rangle$ includes a corrective term for the case of $\phi>0$ that reduces the number of photons in the resonator. To derive the modified relation, we follow the method of Ref.~\onlinecite{SchneiderThesis}. Under a steady state condition, the average energy stored within the resonator is $\langle n_r \rangle h f_r = P_{loss} Q_i/2\pi f_r$, where $h$ is Planck's constant and $P_{loss}$ is the energy dissipated per resonance cycle. The dissipated power is the difference between the chip input power $P_{chip}$ and the transmitted and reflected powers, which are respectively proportional to the scattering parameters $|S_{21}|^2$ and $|S_{11}|^2$. The dissipated power is then $P_{loss} = P_{chip}(1-|S_{21}|^2-|S_{11}|^2)$.

The expression for $|S_{21}|^2$ and $|S_{11}|^2$ can be simplified by assuming an ideal microwave background and evaluating only at $f=f_r$. In this case and using Eq.~\ref{resEq}, the transmission coefficient for power is 
\begin{equation}
|S_{21}|^2 = |Q_c|^{-2}\big(|Q_c|^2 + Q_l^2 - 2 Q_l |Q_c| \cos (\phi)\big).
\end{equation}
On the other hand, $|S_{11}|^2$ is independent \cite{BraumuellerThesis} of $\phi$ with $|S_{11}|^2 = Q_l^2/|Q_c|^2$ at $f=f_r$. Remembering that $Q_i^{-1} = Q_l^{-1} - (|Q_c|\cos(\phi))^{-1}$, the average photon number can then be calculated as
\begin{equation}
\langle n_r \rangle = \frac{P_{chip}}{2\pi h f_r^2}\frac{2 Q_l^2 \cos(\phi)}{|Q_c|}.
\label{eq:nr}
\end{equation}
This expression is consistent with the case of an ideal notch-type resonator with $\phi=0$. Interestingly, this implies that when $Q_c$ is purely imaginary, i.e. $\phi=\pi/2$, then the average photon number goes to zero, however such a scenario is unphysical as it would require an infinite circuit capacitance \cite{Khalil2012}.

\begin{figure*}
\includegraphics[scale = 0.50]{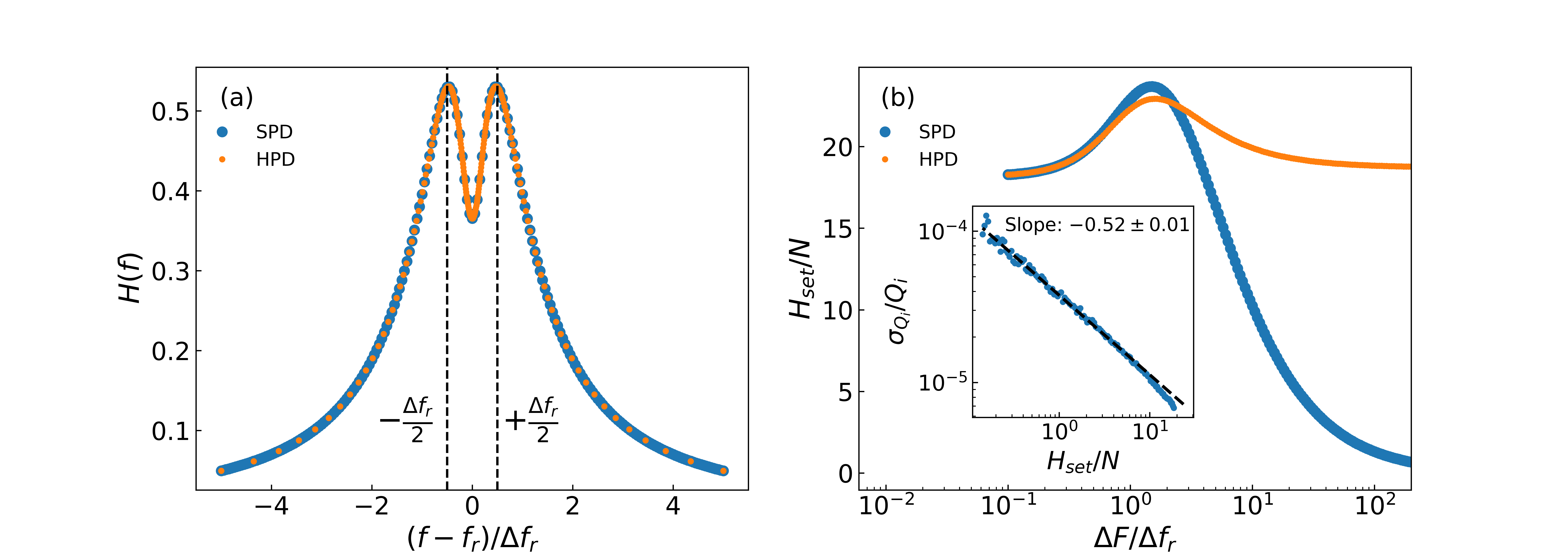}
\caption{(a) The Shannon entropy $H$ as a function of frequency $f$ for a resonance signal with an ideal background and $Q_i= 50000$, $|Q_c|=10000$, $f_r = 5$~GHz, $\phi = 0$, and $N = 201$. The spectrum of $H$ has two peaks at $\pm 0.5 \Delta f_r$ (dashed lines), which correspond to the half-maximum points of the Lorentzian distribution of the resonance signal. These points also correspond to opposite points on the complex resonance circle and are separated by the full diameter of the circle. (b) The normalized information content or information density $H_{set}/N$ for SPD and HPD data sets with $N=10001$. $H_{set}/N$ possesses a peak at $\Delta F/\Delta f_r\sim 1.5$, implying that a half circle measurement is maximally efficient for information content. The inset shows that the fitting error $\sigma_{Q_i}/Q_i$ has an approximate inverse square-root dependence on $H_{set}/N$ with higher amounts of information content corresponding to lower levels of fit uncertainty. \label{fig:entropy}}
\end{figure*}

\section{Appendix F: The Informational Density of a Resonance Data Set}

The purpose of a measurement is to gain information about the system under probe. Depending on the measurement details, a data set may contain more or less information about the system. For instance, even for very large $N$, a measurement centered at $f = 4$~GHz will contain very little information about a narrow resonance signal at $f_r=5$~GHz. In this line of reasoning, one may expect that data points measured further and further away from $f_r$ will contribute increasingly diminishing amounts of information about the resonance signal itself. As noted in the main text, fitting errors for the SPD are minimal for $\Delta F = \Delta f_r$ and increase steadily as data points are dispersed further and further away from $f_r$ as $\Delta F$ increases. Therefore, there is an intuitive relation between fit uncertainty and the information within a data set. Indeed, by calculating the information content of a resonance data set, we will show that such a relation does seem to exist. Furthermore, we will show that for wide ranges the HPD contains more information over the SPD, which helps explain why the HPD returns smaller fit errors.

The resonator system is probed by sending a known electromagnetic signal through a coupled transmission line. The electromagnetic signal sent to the chip by the VNA is a wavepacket consisting of $\langle n_{ph} \rangle$ photons at frequency $f$ and phase $\phi$. From a semi-classical standpoint, a change in the number of photons can occur via an interaction with the system under probe, i.e. the resonator. Should no change to the input signal occur, then no information is gained from the measurement. On the other hand, should the number of photons, their frequency, or their phase change, then such a change would indicate that new information has been gained by performing the measurement. This system can be mapped onto a binary communication system in which a constant bit undergoes a random bit-flip process during transmission through a noisy communication channel, as shown in Fig.~\ref{fig:entropy}(a). If the photon number is considered in isolation (i.e. no frequency or phase shifts occur in the signal), then the informational yield per photon can be parameterized by the Shannon entropy \cite{Shannon1948} $H(p_r) = -p_r \log_2(p_r)$, where $p_r$ is the probability of a singular photon leaving the wavepacket to enter the resonator (or a singular input bit is flipped). For a transmission measurement, the VNA measures the ratio of output to input voltages $V_{output}/V_{input}$, which is equivalent to the ratio of the change in the number of photons $\langle n_{ph}' \rangle/\langle n_{ph} \rangle$, where $\langle n_{ph}' \rangle$ is the number of photons after transmission through the circuit. Therefore, under these conditions, a VNA transmission measurement is a measurement of the probability $p_r$, which is frequency dependent with a Lorentzian distribution $p_r(f_i) = |1-S_{21}|^2$.

$H(p_r)$ is a metric of how unpredictable a measurement's outcome will be \cite{Jaynes1957}, and therefore measurements with larger $H(p_r)$ yield more information about the system. Since $p_r$, and therefore $H(p_r)$, is frequency dependent, not all points in a VNA measurement yield equal amounts of information about the resonance signal. Figure~\ref{fig:entropy}(a) shows the $H(f)$ distribution for a slightly overcoupled resonance signal with an ideal background and $Q_i= 50000$, $|Q_c|=10000$, $f_r = 5$~GHz, and $\phi = 0$. Near $f_r$ are two sharp peaks in $H(f)$, where each is centered at $\pm 0.5 \Delta f_r$. At $f_r$ itself is a dip with a magnitude that depends on the quality factors. $H(f)$ is a metric of the "surprise" that a photon will be absorbed by the resonator. In this case, far from $f_r$ there is no surprise that a photon is not absorbed, hence the diminishing entropy far from $f_r$. On the other hand, at $f_r$ there is little surprise that a photon will be absorbed into the resonator because this is where the peak absorption is. Indeed, for very high quality factor resonators, where the probability of absorption trends toward 1 at $f_r$, the information entropy at $f_r$ goes to zero as there will be no surprise for absorption (not shown). The two peaks at $\pm 0.5 \Delta f_r$ occur because at these values there is an equal $50\%$ probability that a photon will be or not be absorbed into the resonator. Thus the result of a single photon measurement is unpredictable, and therefore these two frequency points have a maximal entropy. 

For a measurement, each data point at frequency $f_i$ yields a small portion of information corresponding to the entropy $H(f_i) = -p_r(f_i) \log_2(p_r(f_i))$. The total information yield in the data set is then the sum of information yield at each measurement frequency $H_{set} = \sum H(f_i)$. We normalize $H_{set}$ by the number of data points to find a metric for the total information density $H_{set}/N$. Figure~\ref{fig:entropy}(b) shows $H_{set}/N$ for a series of simulated SPD and HPD resonance data with varying $\Delta F/\Delta f_r$. Both point distributions have a peak in information density near $\Delta F/\Delta f_r\sim 1.5$, which corresponds to a measurement of a little more than half a resonance circle. Interestingly, the SPD has a slightly higher peak in entropy compared to the HPD, however this behavior is seemingly masked by measurement noise in our results in Fig.~\ref{fig:hpd}. In any case, as $\Delta F$ increases away from $\Delta f_r$, $H_{set}/N$ decreases rapidly for the SPD but only slightly for the HPD. Therefore, the HPD maintains a constant and overall higher information yield at large $\Delta F$. 

Since the entropy is a metric of the uncertainty in the system \cite{Jaynes1957}, data points with the largest amounts of uncertainty yield the most information when measured. Therefore, increasing $H_{set}$ of a measurement should yield more information, which translates into better fit accuracy. Indeed, we find that $H_{set}$ corresponds well with the fit uncertainty. The inset of Fig.~\ref{fig:entropy}(b) shows the dependence of $\sigma_{Q_i}/Q_i$ with respect to $H_{set}/N$. There is an approximate inverse square-root relationship $\sigma_{Q_i}/Q_i \propto (H_{set}/N)^{-1/2}$, and as $H_{set}/N$ decreases for both $\Delta F<\Delta f_r$ and $\Delta F>\Delta f_r$, $\sigma_{Q_i}/Q_i$ increases. Therefore, there is a good correspondence between the total information entropy of a VNA scan and the minimal uncertainty that can be achieved with a circle fit algorithm. One benefit of the HPD is that it better preserves the maximal entropy of the data set compared to the SPD. We note, however, that the analysis above assumes that there is no phase or frequency shift in the photons after traveling through the circuit. When this assumption is false, i.e. when $\phi \neq 0$, then $p_r \neq |1-S_{21}|^2$, since $|S_{21}(f_i)| > 1$ at some $f_i$ due to rotation of the circle. Additionally, sources of measurement error, such as additional photons being added or removed by thermal fluctuations in the measurement circuit, are also neglected. Instead, a more complete analysis which takes into consideration measurement noise and the information gained by a phase shift is required. Such an analysis may be an avenue for future research in resonance measurement theory.

\clearpage
\bibliography{refs}

\providecommand{\noopsort}[1]{}\providecommand{\singleletter}[1]{#1}%
\begin{thebibliography}{50}%
\makeatletter
\providecommand \@ifxundefined [1]{%
 \@ifx{#1\undefined}
}%
\providecommand \@ifnum [1]{%
 \ifnum #1\expandafter \@firstoftwo
 \else \expandafter \@secondoftwo
 \fi
}%
\providecommand \@ifx [1]{%
 \ifx #1\expandafter \@firstoftwo
 \else \expandafter \@secondoftwo
 \fi
}%
\providecommand \natexlab [1]{#1}%
\providecommand \enquote  [1]{``#1''}%
\providecommand \bibnamefont  [1]{#1}%
\providecommand \bibfnamefont [1]{#1}%
\providecommand \citenamefont [1]{#1}%
\providecommand \href@noop [0]{\@secondoftwo}%
\providecommand \href [0]{\begingroup \@sanitize@url \@href}%
\providecommand \@href[1]{\@@startlink{#1}\@@href}%
\providecommand \@@href[1]{\endgroup#1\@@endlink}%
\providecommand \@sanitize@url [0]{\catcode `\\12\catcode `\$12\catcode
  `\&12\catcode `\#12\catcode `\^12\catcode `\_12\catcode `\%12\relax}%
\providecommand \@@startlink[1]{}%
\providecommand \@@endlink[0]{}%
\providecommand \url  [0]{\begingroup\@sanitize@url \@url }%
\providecommand \@url [1]{\endgroup\@href {#1}{\urlprefix }}%
\providecommand \urlprefix  [0]{URL }%
\providecommand \Eprint [0]{\href }%
\providecommand \doibase [0]{https://doi.org/}%
\providecommand \selectlanguage [0]{\@gobble}%
\providecommand \bibinfo  [0]{\@secondoftwo}%
\providecommand \bibfield  [0]{\@secondoftwo}%
\providecommand \translation [1]{[#1]}%
\providecommand \BibitemOpen [0]{}%
\providecommand \bibitemStop [0]{}%
\providecommand \bibitemNoStop [0]{.\EOS\space}%
\providecommand \EOS [0]{\spacefactor3000\relax}%
\providecommand \BibitemShut  [1]{\csname bibitem#1\endcsname}%
\let\auto@bib@innerbib\@empty
\bibitem [{\citenamefont {Place}\ \emph {et~al.}(2021)\citenamefont {Place},
  \citenamefont {Rodgers}, \citenamefont {Mundada}, \citenamefont {Smitham},
  \citenamefont {Fitzpatrick}, \citenamefont {Leng}, \citenamefont {Premkumar},
  \citenamefont {Bryon}, \citenamefont {Vrajitoarea}, \citenamefont {Sussman},
  \citenamefont {Cheng}, \citenamefont {Madhavan}, \citenamefont {Babla},
  \citenamefont {Le}, \citenamefont {Gang}, \citenamefont {J{\"a}ck},
  \citenamefont {Gyenis}, \citenamefont {Yao}, \citenamefont {Cava},
  \citenamefont {de~Leon},\ and\ \citenamefont {Houck}}]{Place2021}%
  \BibitemOpen
  \bibfield  {author} {\bibinfo {author} {\bibfnamefont {A.~P.~M.}\
  \bibnamefont {Place}}, \bibinfo {author} {\bibfnamefont {L.~V.~H.}\
  \bibnamefont {Rodgers}}, \bibinfo {author} {\bibfnamefont {P.}~\bibnamefont
  {Mundada}}, \bibinfo {author} {\bibfnamefont {B.~M.}\ \bibnamefont
  {Smitham}}, \bibinfo {author} {\bibfnamefont {M.}~\bibnamefont
  {Fitzpatrick}}, \bibinfo {author} {\bibfnamefont {Z.}~\bibnamefont {Leng}},
  \bibinfo {author} {\bibfnamefont {A.}~\bibnamefont {Premkumar}}, \bibinfo
  {author} {\bibfnamefont {J.}~\bibnamefont {Bryon}}, \bibinfo {author}
  {\bibfnamefont {A.}~\bibnamefont {Vrajitoarea}}, \bibinfo {author}
  {\bibfnamefont {S.}~\bibnamefont {Sussman}}, \bibinfo {author} {\bibfnamefont
  {G.}~\bibnamefont {Cheng}}, \bibinfo {author} {\bibfnamefont
  {T.}~\bibnamefont {Madhavan}}, \bibinfo {author} {\bibfnamefont {H.~K.}\
  \bibnamefont {Babla}}, \bibinfo {author} {\bibfnamefont {X.~H.}\ \bibnamefont
  {Le}}, \bibinfo {author} {\bibfnamefont {Y.}~\bibnamefont {Gang}}, \bibinfo
  {author} {\bibfnamefont {B.}~\bibnamefont {J{\"a}ck}}, \bibinfo {author}
  {\bibfnamefont {A.}~\bibnamefont {Gyenis}}, \bibinfo {author} {\bibfnamefont
  {N.}~\bibnamefont {Yao}}, \bibinfo {author} {\bibfnamefont {R.~J.}\
  \bibnamefont {Cava}}, \bibinfo {author} {\bibfnamefont {N.~P.}\ \bibnamefont
  {de~Leon}},\ and\ \bibinfo {author} {\bibfnamefont {A.~A.}\ \bibnamefont
  {Houck}},\ }\bibfield  {title} {\bibinfo {title} {New material platform for
  superconducting transmon qubits with coherence times exceeding 0.3
  milliseconds},\ }\href {https://doi.org/10.1038/s41467-021-22030-5}
  {\bibfield  {journal} {\bibinfo  {journal} {Nature Communications}\ }\textbf
  {\bibinfo {volume} {12}},\ \bibinfo {pages} {1779} (\bibinfo {year}
  {2021})}\BibitemShut {NoStop}%
\bibitem [{\citenamefont {Murray}(2021)}]{Murray2021}%
  \BibitemOpen
  \bibfield  {author} {\bibinfo {author} {\bibfnamefont {C.~E.}\ \bibnamefont
  {Murray}},\ }\bibfield  {title} {\bibinfo {title} {Material matters in
  superconducting qubits},\ }\href
  {https://doi.org/https://doi.org/10.1016/j.mser.2021.100646} {\bibfield
  {journal} {\bibinfo  {journal} {Materials Science and Engineering: R:
  Reports}\ }\textbf {\bibinfo {volume} {146}},\ \bibinfo {pages} {100646}
  (\bibinfo {year} {2021})}\BibitemShut {NoStop}%
\bibitem [{\citenamefont {Wang}\ \emph {et~al.}(2022)\citenamefont {Wang},
  \citenamefont {Li}, \citenamefont {Xu}, \citenamefont {Li}, \citenamefont
  {Wang}, \citenamefont {Yang}, \citenamefont {Mi}, \citenamefont {Liang},
  \citenamefont {Su}, \citenamefont {Yang}, \citenamefont {Wang}, \citenamefont
  {Wang}, \citenamefont {Li}, \citenamefont {Chen}, \citenamefont {Li},
  \citenamefont {Linghu}, \citenamefont {Han}, \citenamefont {Zhang},
  \citenamefont {Feng}, \citenamefont {Song}, \citenamefont {Ma}, \citenamefont
  {Zhang}, \citenamefont {Wang}, \citenamefont {Zhao}, \citenamefont {Liu},
  \citenamefont {Xue}, \citenamefont {Jin},\ and\ \citenamefont
  {Yu}}]{Wang2022}%
  \BibitemOpen
  \bibfield  {author} {\bibinfo {author} {\bibfnamefont {C.}~\bibnamefont
  {Wang}}, \bibinfo {author} {\bibfnamefont {X.}~\bibnamefont {Li}}, \bibinfo
  {author} {\bibfnamefont {H.}~\bibnamefont {Xu}}, \bibinfo {author}
  {\bibfnamefont {Z.}~\bibnamefont {Li}}, \bibinfo {author} {\bibfnamefont
  {J.}~\bibnamefont {Wang}}, \bibinfo {author} {\bibfnamefont {Z.}~\bibnamefont
  {Yang}}, \bibinfo {author} {\bibfnamefont {Z.}~\bibnamefont {Mi}}, \bibinfo
  {author} {\bibfnamefont {X.}~\bibnamefont {Liang}}, \bibinfo {author}
  {\bibfnamefont {T.}~\bibnamefont {Su}}, \bibinfo {author} {\bibfnamefont
  {C.}~\bibnamefont {Yang}}, \bibinfo {author} {\bibfnamefont {G.}~\bibnamefont
  {Wang}}, \bibinfo {author} {\bibfnamefont {W.}~\bibnamefont {Wang}}, \bibinfo
  {author} {\bibfnamefont {Y.}~\bibnamefont {Li}}, \bibinfo {author}
  {\bibfnamefont {M.}~\bibnamefont {Chen}}, \bibinfo {author} {\bibfnamefont
  {C.}~\bibnamefont {Li}}, \bibinfo {author} {\bibfnamefont {K.}~\bibnamefont
  {Linghu}}, \bibinfo {author} {\bibfnamefont {J.}~\bibnamefont {Han}},
  \bibinfo {author} {\bibfnamefont {Y.}~\bibnamefont {Zhang}}, \bibinfo
  {author} {\bibfnamefont {Y.}~\bibnamefont {Feng}}, \bibinfo {author}
  {\bibfnamefont {Y.}~\bibnamefont {Song}}, \bibinfo {author} {\bibfnamefont
  {T.}~\bibnamefont {Ma}}, \bibinfo {author} {\bibfnamefont {J.}~\bibnamefont
  {Zhang}}, \bibinfo {author} {\bibfnamefont {R.}~\bibnamefont {Wang}},
  \bibinfo {author} {\bibfnamefont {P.}~\bibnamefont {Zhao}}, \bibinfo {author}
  {\bibfnamefont {W.}~\bibnamefont {Liu}}, \bibinfo {author} {\bibfnamefont
  {G.}~\bibnamefont {Xue}}, \bibinfo {author} {\bibfnamefont {Y.}~\bibnamefont
  {Jin}},\ and\ \bibinfo {author} {\bibfnamefont {H.}~\bibnamefont {Yu}},\
  }\bibfield  {title} {\bibinfo {title} {Towards practical quantum computers:
  transmon qubit with a lifetime approaching 0.5 milliseconds},\ }\href
  {https://doi.org/10.1038/s41534-021-00510-2} {\bibfield  {journal} {\bibinfo
  {journal} {npj Quantum Information}\ }\textbf {\bibinfo {volume} {8}},\
  \bibinfo {pages} {3} (\bibinfo {year} {2022})}\BibitemShut {NoStop}%
\bibitem [{\citenamefont {Iaia}\ \emph {et~al.}(2022)\citenamefont {Iaia},
  \citenamefont {Ku}, \citenamefont {Ballard}, \citenamefont {Larson},
  \citenamefont {Yelton}, \citenamefont {Liu}, \citenamefont {Patel},
  \citenamefont {McDermott},\ and\ \citenamefont {Plourde}}]{Iaia2022}%
  \BibitemOpen
  \bibfield  {author} {\bibinfo {author} {\bibfnamefont {V.}~\bibnamefont
  {Iaia}}, \bibinfo {author} {\bibfnamefont {J.}~\bibnamefont {Ku}}, \bibinfo
  {author} {\bibfnamefont {A.}~\bibnamefont {Ballard}}, \bibinfo {author}
  {\bibfnamefont {C.~P.}\ \bibnamefont {Larson}}, \bibinfo {author}
  {\bibfnamefont {E.}~\bibnamefont {Yelton}}, \bibinfo {author} {\bibfnamefont
  {C.~H.}\ \bibnamefont {Liu}}, \bibinfo {author} {\bibfnamefont
  {S.}~\bibnamefont {Patel}}, \bibinfo {author} {\bibfnamefont
  {R.}~\bibnamefont {McDermott}},\ and\ \bibinfo {author} {\bibfnamefont
  {B.~L.~T.}\ \bibnamefont {Plourde}},\ }\bibfield  {title} {\bibinfo {title}
  {Phonon downconversion to suppress correlated errors in superconducting
  qubits},\ }\href {https://doi.org/10.1038/s41467-022-33997-0} {\bibfield
  {journal} {\bibinfo  {journal} {Nature Communications}\ }\textbf {\bibinfo
  {volume} {13}},\ \bibinfo {pages} {6425} (\bibinfo {year}
  {2022})}\BibitemShut {NoStop}%
\bibitem [{\citenamefont {Verjauw}\ \emph {et~al.}(2022)\citenamefont
  {Verjauw}, \citenamefont {Acharya}, \citenamefont {Van~Damme}, \citenamefont
  {Ivanov}, \citenamefont {Lozano}, \citenamefont {Mohiyaddin}, \citenamefont
  {Wan}, \citenamefont {Jussot}, \citenamefont {Vadiraj}, \citenamefont
  {Mongillo}, \citenamefont {Heyns}, \citenamefont {Radu}, \citenamefont
  {Govoreanu},\ and\ \citenamefont {Poto{\v{c}}nik}}]{Verjauw2022}%
  \BibitemOpen
  \bibfield  {author} {\bibinfo {author} {\bibfnamefont {J.}~\bibnamefont
  {Verjauw}}, \bibinfo {author} {\bibfnamefont {R.}~\bibnamefont {Acharya}},
  \bibinfo {author} {\bibfnamefont {J.}~\bibnamefont {Van~Damme}}, \bibinfo
  {author} {\bibfnamefont {T.}~\bibnamefont {Ivanov}}, \bibinfo {author}
  {\bibfnamefont {D.~P.}\ \bibnamefont {Lozano}}, \bibinfo {author}
  {\bibfnamefont {F.~A.}\ \bibnamefont {Mohiyaddin}}, \bibinfo {author}
  {\bibfnamefont {D.}~\bibnamefont {Wan}}, \bibinfo {author} {\bibfnamefont
  {J.}~\bibnamefont {Jussot}}, \bibinfo {author} {\bibfnamefont {A.~M.}\
  \bibnamefont {Vadiraj}}, \bibinfo {author} {\bibfnamefont {M.}~\bibnamefont
  {Mongillo}}, \bibinfo {author} {\bibfnamefont {M.}~\bibnamefont {Heyns}},
  \bibinfo {author} {\bibfnamefont {I.}~\bibnamefont {Radu}}, \bibinfo {author}
  {\bibfnamefont {B.}~\bibnamefont {Govoreanu}},\ and\ \bibinfo {author}
  {\bibfnamefont {A.}~\bibnamefont {Poto{\v{c}}nik}},\ }\bibfield  {title}
  {\bibinfo {title} {Path toward manufacturable superconducting qubits with
  relaxation times exceeding 0.1{\thinspace}ms},\ }\href
  {https://doi.org/10.1038/s41534-022-00600-9} {\bibfield  {journal} {\bibinfo
  {journal} {npj Quantum Information}\ }\textbf {\bibinfo {volume} {8}},\
  \bibinfo {pages} {93} (\bibinfo {year} {2022})}\BibitemShut {NoStop}%
\bibitem [{\citenamefont {Krantz}\ \emph {et~al.}(2019)\citenamefont {Krantz},
  \citenamefont {Kjaergaard}, \citenamefont {Yan}, \citenamefont {Orlando},
  \citenamefont {Gustavsson},\ and\ \citenamefont {Oliver}}]{Krantz2019}%
  \BibitemOpen
  \bibfield  {author} {\bibinfo {author} {\bibfnamefont {P.}~\bibnamefont
  {Krantz}}, \bibinfo {author} {\bibfnamefont {M.}~\bibnamefont {Kjaergaard}},
  \bibinfo {author} {\bibfnamefont {F.}~\bibnamefont {Yan}}, \bibinfo {author}
  {\bibfnamefont {T.~P.}\ \bibnamefont {Orlando}}, \bibinfo {author}
  {\bibfnamefont {S.}~\bibnamefont {Gustavsson}},\ and\ \bibinfo {author}
  {\bibfnamefont {W.~D.}\ \bibnamefont {Oliver}},\ }\bibfield  {title}
  {\bibinfo {title} {A quantum engineer's guide to superconducting qubits},\
  }\href {https://doi.org/10.1063/1.5089550} {\bibfield  {journal} {\bibinfo
  {journal} {Applied Physics Reviews}\ }\textbf {\bibinfo {volume} {6}},\
  \bibinfo {pages} {021318} (\bibinfo {year} {2019})},\ \Eprint
  {https://arxiv.org/abs/https://doi.org/10.1063/1.5089550}
  {https://doi.org/10.1063/1.5089550} \BibitemShut {NoStop}%
\bibitem [{\citenamefont {Rossi}\ \emph {et~al.}(2021)\citenamefont {Rossi},
  \citenamefont {Baity}, \citenamefont {Schäfer},\ and\ \citenamefont
  {Weides}}]{Rossi2021}%
  \BibitemOpen
  \bibfield  {author} {\bibinfo {author} {\bibfnamefont {A.}~\bibnamefont
  {Rossi}}, \bibinfo {author} {\bibfnamefont {P.~G.}\ \bibnamefont {Baity}},
  \bibinfo {author} {\bibfnamefont {V.~M.}\ \bibnamefont {Schäfer}},\ and\
  \bibinfo {author} {\bibfnamefont {M.}~\bibnamefont {Weides}},\ }\bibfield
  {title} {\bibinfo {title} {Quantum computing hardware in the cloud: Should a
  computational chemist care?},\ }\href
  {https://doi.org/https://doi.org/10.1002/qua.26688} {\bibfield  {journal}
  {\bibinfo  {journal} {International Journal of Quantum Chemistry}\ }\textbf
  {\bibinfo {volume} {121}},\ \bibinfo {pages} {e26688} (\bibinfo {year}
  {2021})},\ \Eprint
  {https://arxiv.org/abs/https://onlinelibrary.wiley.com/doi/pdf/10.1002/qua.26688}
  {https://onlinelibrary.wiley.com/doi/pdf/10.1002/qua.26688} \BibitemShut
  {NoStop}%
\bibitem [{\citenamefont {Megrant}\ \emph {et~al.}(2012)\citenamefont
  {Megrant}, \citenamefont {Neill}, \citenamefont {Barends}, \citenamefont
  {Chiaro}, \citenamefont {Chen}, \citenamefont {Feigl}, \citenamefont {Kelly},
  \citenamefont {Lucero}, \citenamefont {Mariantoni}, \citenamefont
  {O’Malley}, \citenamefont {Sank}, \citenamefont {Vainsencher},
  \citenamefont {Wenner}, \citenamefont {White}, \citenamefont {Yin},
  \citenamefont {Zhao}, \citenamefont {Palmstrøm}, \citenamefont {Martinis},\
  and\ \citenamefont {Cleland}}]{Megrant2012}%
  \BibitemOpen
  \bibfield  {author} {\bibinfo {author} {\bibfnamefont {A.}~\bibnamefont
  {Megrant}}, \bibinfo {author} {\bibfnamefont {C.}~\bibnamefont {Neill}},
  \bibinfo {author} {\bibfnamefont {R.}~\bibnamefont {Barends}}, \bibinfo
  {author} {\bibfnamefont {B.}~\bibnamefont {Chiaro}}, \bibinfo {author}
  {\bibfnamefont {Y.}~\bibnamefont {Chen}}, \bibinfo {author} {\bibfnamefont
  {L.}~\bibnamefont {Feigl}}, \bibinfo {author} {\bibfnamefont
  {J.}~\bibnamefont {Kelly}}, \bibinfo {author} {\bibfnamefont
  {E.}~\bibnamefont {Lucero}}, \bibinfo {author} {\bibfnamefont
  {M.}~\bibnamefont {Mariantoni}}, \bibinfo {author} {\bibfnamefont {P.~J.~J.}\
  \bibnamefont {O’Malley}}, \bibinfo {author} {\bibfnamefont
  {D.}~\bibnamefont {Sank}}, \bibinfo {author} {\bibfnamefont {A.}~\bibnamefont
  {Vainsencher}}, \bibinfo {author} {\bibfnamefont {J.}~\bibnamefont {Wenner}},
  \bibinfo {author} {\bibfnamefont {T.~C.}\ \bibnamefont {White}}, \bibinfo
  {author} {\bibfnamefont {Y.}~\bibnamefont {Yin}}, \bibinfo {author}
  {\bibfnamefont {J.}~\bibnamefont {Zhao}}, \bibinfo {author} {\bibfnamefont
  {C.~J.}\ \bibnamefont {Palmstrøm}}, \bibinfo {author} {\bibfnamefont
  {J.~M.}\ \bibnamefont {Martinis}},\ and\ \bibinfo {author} {\bibfnamefont
  {A.~N.}\ \bibnamefont {Cleland}},\ }\bibfield  {title} {\bibinfo {title}
  {Planar superconducting resonators with internal quality factors above one
  million},\ }\href {https://doi.org/10.1063/1.3693409} {\bibfield  {journal}
  {\bibinfo  {journal} {Applied Physics Letters}\ }\textbf {\bibinfo {volume}
  {100}},\ \bibinfo {pages} {113510} (\bibinfo {year} {2012})},\ \Eprint
  {https://arxiv.org/abs/https://doi.org/10.1063/1.3693409}
  {https://doi.org/10.1063/1.3693409} \BibitemShut {NoStop}%
\bibitem [{\citenamefont {Bruno}\ \emph {et~al.}(2015)\citenamefont {Bruno},
  \citenamefont {de~Lange}, \citenamefont {Asaad}, \citenamefont {van~der
  Enden}, \citenamefont {Langford},\ and\ \citenamefont {DiCarlo}}]{Bruno2015}%
  \BibitemOpen
  \bibfield  {author} {\bibinfo {author} {\bibfnamefont {A.}~\bibnamefont
  {Bruno}}, \bibinfo {author} {\bibfnamefont {G.}~\bibnamefont {de~Lange}},
  \bibinfo {author} {\bibfnamefont {S.}~\bibnamefont {Asaad}}, \bibinfo
  {author} {\bibfnamefont {K.~L.}\ \bibnamefont {van~der Enden}}, \bibinfo
  {author} {\bibfnamefont {N.~K.}\ \bibnamefont {Langford}},\ and\ \bibinfo
  {author} {\bibfnamefont {L.}~\bibnamefont {DiCarlo}},\ }\bibfield  {title}
  {\bibinfo {title} {Reducing intrinsic loss in superconducting resonators by
  surface treatment and deep etching of silicon substrates},\ }\href
  {https://doi.org/10.1063/1.4919761} {\bibfield  {journal} {\bibinfo
  {journal} {Applied Physics Letters}\ }\textbf {\bibinfo {volume} {106}},\
  \bibinfo {pages} {182601} (\bibinfo {year} {2015})},\ \Eprint
  {https://arxiv.org/abs/https://doi.org/10.1063/1.4919761}
  {https://doi.org/10.1063/1.4919761} \BibitemShut {NoStop}%
\bibitem [{\citenamefont {Dunsworth}\ \emph {et~al.}(2017)\citenamefont
  {Dunsworth}, \citenamefont {Megrant}, \citenamefont {Quintana}, \citenamefont
  {Chen}, \citenamefont {Barends}, \citenamefont {Burkett}, \citenamefont
  {Foxen}, \citenamefont {Chen}, \citenamefont {Chiaro}, \citenamefont
  {Fowler}, \citenamefont {Graff}, \citenamefont {Jeffrey}, \citenamefont
  {Kelly}, \citenamefont {Lucero}, \citenamefont {Mutus}, \citenamefont
  {Neeley}, \citenamefont {Neill}, \citenamefont {Roushan}, \citenamefont
  {Sank}, \citenamefont {Vainsencher}, \citenamefont {Wenner}, \citenamefont
  {White},\ and\ \citenamefont {Martinis}}]{Dunsworth2017}%
  \BibitemOpen
  \bibfield  {author} {\bibinfo {author} {\bibfnamefont {A.}~\bibnamefont
  {Dunsworth}}, \bibinfo {author} {\bibfnamefont {A.}~\bibnamefont {Megrant}},
  \bibinfo {author} {\bibfnamefont {C.}~\bibnamefont {Quintana}}, \bibinfo
  {author} {\bibfnamefont {Z.}~\bibnamefont {Chen}}, \bibinfo {author}
  {\bibfnamefont {R.}~\bibnamefont {Barends}}, \bibinfo {author} {\bibfnamefont
  {B.}~\bibnamefont {Burkett}}, \bibinfo {author} {\bibfnamefont
  {B.}~\bibnamefont {Foxen}}, \bibinfo {author} {\bibfnamefont
  {Y.}~\bibnamefont {Chen}}, \bibinfo {author} {\bibfnamefont {B.}~\bibnamefont
  {Chiaro}}, \bibinfo {author} {\bibfnamefont {A.}~\bibnamefont {Fowler}},
  \bibinfo {author} {\bibfnamefont {R.}~\bibnamefont {Graff}}, \bibinfo
  {author} {\bibfnamefont {E.}~\bibnamefont {Jeffrey}}, \bibinfo {author}
  {\bibfnamefont {J.}~\bibnamefont {Kelly}}, \bibinfo {author} {\bibfnamefont
  {E.}~\bibnamefont {Lucero}}, \bibinfo {author} {\bibfnamefont {J.~Y.}\
  \bibnamefont {Mutus}}, \bibinfo {author} {\bibfnamefont {M.}~\bibnamefont
  {Neeley}}, \bibinfo {author} {\bibfnamefont {C.}~\bibnamefont {Neill}},
  \bibinfo {author} {\bibfnamefont {P.}~\bibnamefont {Roushan}}, \bibinfo
  {author} {\bibfnamefont {D.}~\bibnamefont {Sank}}, \bibinfo {author}
  {\bibfnamefont {A.}~\bibnamefont {Vainsencher}}, \bibinfo {author}
  {\bibfnamefont {J.}~\bibnamefont {Wenner}}, \bibinfo {author} {\bibfnamefont
  {T.~C.}\ \bibnamefont {White}},\ and\ \bibinfo {author} {\bibfnamefont
  {J.~M.}\ \bibnamefont {Martinis}},\ }\bibfield  {title} {\bibinfo {title}
  {Characterization and reduction of capacitive loss induced by sub-micron
  josephson junction fabrication in superconducting qubits},\ }\href
  {https://doi.org/10.1063/1.4993577} {\bibfield  {journal} {\bibinfo
  {journal} {Applied Physics Letters}\ }\textbf {\bibinfo {volume} {111}},\
  \bibinfo {pages} {022601} (\bibinfo {year} {2017})},\ \Eprint
  {https://arxiv.org/abs/https://doi.org/10.1063/1.4993577}
  {https://doi.org/10.1063/1.4993577} \BibitemShut {NoStop}%
\bibitem [{\citenamefont {Henriques}\ \emph {et~al.}(2019)\citenamefont
  {Henriques}, \citenamefont {Valenti}, \citenamefont {Charpentier},
  \citenamefont {Lagoin}, \citenamefont {Gouriou}, \citenamefont {Martínez},
  \citenamefont {Cardani}, \citenamefont {Vignati}, \citenamefont {Grünhaupt},
  \citenamefont {Gusenkova}, \citenamefont {Ferrero}, \citenamefont {Skacel},
  \citenamefont {Wernsdorfer}, \citenamefont {Ustinov}, \citenamefont
  {Catelani}, \citenamefont {Sander},\ and\ \citenamefont
  {Pop}}]{Henriques2019}%
  \BibitemOpen
  \bibfield  {author} {\bibinfo {author} {\bibfnamefont {F.}~\bibnamefont
  {Henriques}}, \bibinfo {author} {\bibfnamefont {F.}~\bibnamefont {Valenti}},
  \bibinfo {author} {\bibfnamefont {T.}~\bibnamefont {Charpentier}}, \bibinfo
  {author} {\bibfnamefont {M.}~\bibnamefont {Lagoin}}, \bibinfo {author}
  {\bibfnamefont {C.}~\bibnamefont {Gouriou}}, \bibinfo {author} {\bibfnamefont
  {M.}~\bibnamefont {Martínez}}, \bibinfo {author} {\bibfnamefont
  {L.}~\bibnamefont {Cardani}}, \bibinfo {author} {\bibfnamefont
  {M.}~\bibnamefont {Vignati}}, \bibinfo {author} {\bibfnamefont
  {L.}~\bibnamefont {Grünhaupt}}, \bibinfo {author} {\bibfnamefont
  {D.}~\bibnamefont {Gusenkova}}, \bibinfo {author} {\bibfnamefont
  {J.}~\bibnamefont {Ferrero}}, \bibinfo {author} {\bibfnamefont {S.~T.}\
  \bibnamefont {Skacel}}, \bibinfo {author} {\bibfnamefont {W.}~\bibnamefont
  {Wernsdorfer}}, \bibinfo {author} {\bibfnamefont {A.~V.}\ \bibnamefont
  {Ustinov}}, \bibinfo {author} {\bibfnamefont {G.}~\bibnamefont {Catelani}},
  \bibinfo {author} {\bibfnamefont {O.}~\bibnamefont {Sander}},\ and\ \bibinfo
  {author} {\bibfnamefont {I.~M.}\ \bibnamefont {Pop}},\ }\bibfield  {title}
  {\bibinfo {title} {Phonon traps reduce the quasiparticle density in
  superconducting circuits},\ }\href {https://doi.org/10.1063/1.5124967}
  {\bibfield  {journal} {\bibinfo  {journal} {Applied Physics Letters}\
  }\textbf {\bibinfo {volume} {115}},\ \bibinfo {pages} {212601} (\bibinfo
  {year} {2019})},\ \Eprint
  {https://arxiv.org/abs/https://doi.org/10.1063/1.5124967}
  {https://doi.org/10.1063/1.5124967} \BibitemShut {NoStop}%
\bibitem [{\citenamefont {McRae}\ \emph {et~al.}(2020)\citenamefont {McRae},
  \citenamefont {Wang}, \citenamefont {Gao}, \citenamefont {Vissers},
  \citenamefont {Brecht}, \citenamefont {Dunsworth}, \citenamefont {Pappas},\
  and\ \citenamefont {Mutus}}]{McRae2020}%
  \BibitemOpen
  \bibfield  {author} {\bibinfo {author} {\bibfnamefont {C.~R.~H.}\
  \bibnamefont {McRae}}, \bibinfo {author} {\bibfnamefont {H.}~\bibnamefont
  {Wang}}, \bibinfo {author} {\bibfnamefont {J.}~\bibnamefont {Gao}}, \bibinfo
  {author} {\bibfnamefont {M.~R.}\ \bibnamefont {Vissers}}, \bibinfo {author}
  {\bibfnamefont {T.}~\bibnamefont {Brecht}}, \bibinfo {author} {\bibfnamefont
  {A.}~\bibnamefont {Dunsworth}}, \bibinfo {author} {\bibfnamefont {D.~P.}\
  \bibnamefont {Pappas}},\ and\ \bibinfo {author} {\bibfnamefont
  {J.}~\bibnamefont {Mutus}},\ }\bibfield  {title} {\bibinfo {title} {Materials
  loss measurements using superconducting microwave resonators},\ }\href
  {https://doi.org/10.1063/5.0017378} {\bibfield  {journal} {\bibinfo
  {journal} {Review of Scientific Instruments}\ }\textbf {\bibinfo {volume}
  {91}},\ \bibinfo {pages} {091101} (\bibinfo {year} {2020})},\ \Eprint
  {https://arxiv.org/abs/https://doi.org/10.1063/5.0017378}
  {https://doi.org/10.1063/5.0017378} \BibitemShut {NoStop}%
\bibitem [{\citenamefont {de~Graaf}\ \emph {et~al.}(2020)\citenamefont
  {de~Graaf}, \citenamefont {Faoro}, \citenamefont {Ioffe}, \citenamefont
  {Mahashabde}, \citenamefont {Burnett}, \citenamefont {Lindström},
  \citenamefont {Kubatkin}, \citenamefont {Danilov},\ and\ \citenamefont
  {Tzalenchuk}}]{deGraaf2020}%
  \BibitemOpen
  \bibfield  {author} {\bibinfo {author} {\bibfnamefont {S.~E.}\ \bibnamefont
  {de~Graaf}}, \bibinfo {author} {\bibfnamefont {L.}~\bibnamefont {Faoro}},
  \bibinfo {author} {\bibfnamefont {L.~B.}\ \bibnamefont {Ioffe}}, \bibinfo
  {author} {\bibfnamefont {S.}~\bibnamefont {Mahashabde}}, \bibinfo {author}
  {\bibfnamefont {J.~J.}\ \bibnamefont {Burnett}}, \bibinfo {author}
  {\bibfnamefont {T.}~\bibnamefont {Lindström}}, \bibinfo {author}
  {\bibfnamefont {S.~E.}\ \bibnamefont {Kubatkin}}, \bibinfo {author}
  {\bibfnamefont {A.~V.}\ \bibnamefont {Danilov}},\ and\ \bibinfo {author}
  {\bibfnamefont {A.~Y.}\ \bibnamefont {Tzalenchuk}},\ }\bibfield  {title}
  {\bibinfo {title} {Two-level systems in superconducting quantum devices due
  to trapped quasiparticles},\ }\href {https://doi.org/10.1126/sciadv.abc5055}
  {\bibfield  {journal} {\bibinfo  {journal} {Science Advances}\ }\textbf
  {\bibinfo {volume} {6}},\ \bibinfo {pages} {eabc5055} (\bibinfo {year}
  {2020})},\ \Eprint
  {https://arxiv.org/abs/https://www.science.org/doi/pdf/10.1126/sciadv.abc5055}
  {https://www.science.org/doi/pdf/10.1126/sciadv.abc5055} \BibitemShut
  {NoStop}%
\bibitem [{\citenamefont {Alto\'e}\ \emph {et~al.}(2022)\citenamefont
  {Alto\'e}, \citenamefont {Banerjee}, \citenamefont {Berk}, \citenamefont
  {Hajr}, \citenamefont {Schwartzberg}, \citenamefont {Song}, \citenamefont
  {Alghadeer}, \citenamefont {Aloni}, \citenamefont {Elowson}, \citenamefont
  {Kreikebaum}, \citenamefont {Wong}, \citenamefont {Griffin}, \citenamefont
  {Rao}, \citenamefont {Weber-Bargioni}, \citenamefont {Minor}, \citenamefont
  {Santiago}, \citenamefont {Cabrini}, \citenamefont {Siddiqi},\ and\
  \citenamefont {Ogletree}}]{Altoe2022}%
  \BibitemOpen
  \bibfield  {author} {\bibinfo {author} {\bibfnamefont {M.~V.~P.}\
  \bibnamefont {Alto\'e}}, \bibinfo {author} {\bibfnamefont {A.}~\bibnamefont
  {Banerjee}}, \bibinfo {author} {\bibfnamefont {C.}~\bibnamefont {Berk}},
  \bibinfo {author} {\bibfnamefont {A.}~\bibnamefont {Hajr}}, \bibinfo {author}
  {\bibfnamefont {A.}~\bibnamefont {Schwartzberg}}, \bibinfo {author}
  {\bibfnamefont {C.}~\bibnamefont {Song}}, \bibinfo {author} {\bibfnamefont
  {M.}~\bibnamefont {Alghadeer}}, \bibinfo {author} {\bibfnamefont
  {S.}~\bibnamefont {Aloni}}, \bibinfo {author} {\bibfnamefont {M.~J.}\
  \bibnamefont {Elowson}}, \bibinfo {author} {\bibfnamefont {J.~M.}\
  \bibnamefont {Kreikebaum}}, \bibinfo {author} {\bibfnamefont {E.~K.}\
  \bibnamefont {Wong}}, \bibinfo {author} {\bibfnamefont {S.~M.}\ \bibnamefont
  {Griffin}}, \bibinfo {author} {\bibfnamefont {S.}~\bibnamefont {Rao}},
  \bibinfo {author} {\bibfnamefont {A.}~\bibnamefont {Weber-Bargioni}},
  \bibinfo {author} {\bibfnamefont {A.~M.}\ \bibnamefont {Minor}}, \bibinfo
  {author} {\bibfnamefont {D.~I.}\ \bibnamefont {Santiago}}, \bibinfo {author}
  {\bibfnamefont {S.}~\bibnamefont {Cabrini}}, \bibinfo {author} {\bibfnamefont
  {I.}~\bibnamefont {Siddiqi}},\ and\ \bibinfo {author} {\bibfnamefont {D.~F.}\
  \bibnamefont {Ogletree}},\ }\bibfield  {title} {\bibinfo {title}
  {Localization and mitigation of loss in niobium superconducting circuits},\
  }\href {https://doi.org/10.1103/PRXQuantum.3.020312} {\bibfield  {journal}
  {\bibinfo  {journal} {PRX Quantum}\ }\textbf {\bibinfo {volume} {3}},\
  \bibinfo {pages} {020312} (\bibinfo {year} {2022})}\BibitemShut {NoStop}%
\bibitem [{\citenamefont {Petersan}\ and\ \citenamefont
  {Anlage}(1998)}]{Petersan1998}%
  \BibitemOpen
  \bibfield  {author} {\bibinfo {author} {\bibfnamefont {P.~J.}\ \bibnamefont
  {Petersan}}\ and\ \bibinfo {author} {\bibfnamefont {S.~M.}\ \bibnamefont
  {Anlage}},\ }\bibfield  {title} {\bibinfo {title} {Measurement of resonant
  frequency and quality factor of microwave resonators: Comparison of
  methods},\ }\href {https://doi.org/10.1063/1.368498} {\bibfield  {journal}
  {\bibinfo  {journal} {Journal of Applied Physics}\ }\textbf {\bibinfo
  {volume} {84}},\ \bibinfo {pages} {3392} (\bibinfo {year} {1998})},\ \Eprint
  {https://arxiv.org/abs/https://doi.org/10.1063/1.368498}
  {https://doi.org/10.1063/1.368498} \BibitemShut {NoStop}%
\bibitem [{\citenamefont {McRae}(2022)}]{Mcrae2022}%
  \BibitemOpen
  \bibfield  {author} {\bibinfo {author} {\bibfnamefont {C.~R.~H.}\
  \bibnamefont {McRae}},\ }\bibfield  {title} {\bibinfo {title} {Measurement
  techniques for superconducting microwave resonators towards quantum device
  applications},\ }in\ \href {https://doi.org/10.1109/IMS37962.2022.9865517}
  {\emph {\bibinfo {booktitle} {2022 IEEE/MTT-S International Microwave
  Symposium - IMS 2022}}}\ (\bibinfo {year} {2022})\ pp.\ \bibinfo {pages}
  {230--232}\BibitemShut {NoStop}%
\bibitem [{\citenamefont {Khalil}\ \emph {et~al.}(2012)\citenamefont {Khalil},
  \citenamefont {Stoutimore}, \citenamefont {Wellstood},\ and\ \citenamefont
  {Osborn}}]{Khalil2012}%
  \BibitemOpen
  \bibfield  {author} {\bibinfo {author} {\bibfnamefont {M.~S.}\ \bibnamefont
  {Khalil}}, \bibinfo {author} {\bibfnamefont {M.~J.~A.}\ \bibnamefont
  {Stoutimore}}, \bibinfo {author} {\bibfnamefont {F.~C.}\ \bibnamefont
  {Wellstood}},\ and\ \bibinfo {author} {\bibfnamefont {K.~D.}\ \bibnamefont
  {Osborn}},\ }\bibfield  {title} {\bibinfo {title} {An analysis method for
  asymmetric resonator transmission applied to superconducting devices},\
  }\href {https://doi.org/10.1063/1.3692073} {\bibfield  {journal} {\bibinfo
  {journal} {Journal of Applied Physics}\ }\textbf {\bibinfo {volume} {111}},\
  \bibinfo {pages} {054510} (\bibinfo {year} {2012})},\ \Eprint
  {https://arxiv.org/abs/https://doi.org/10.1063/1.3692073}
  {https://doi.org/10.1063/1.3692073} \BibitemShut {NoStop}%
\bibitem [{\citenamefont {Probst}\ \emph {et~al.}(2015)\citenamefont {Probst},
  \citenamefont {Song}, \citenamefont {Bushev}, \citenamefont {Ustinov},\ and\
  \citenamefont {Weides}}]{Probst2015}%
  \BibitemOpen
  \bibfield  {author} {\bibinfo {author} {\bibfnamefont {S.}~\bibnamefont
  {Probst}}, \bibinfo {author} {\bibfnamefont {F.~B.}\ \bibnamefont {Song}},
  \bibinfo {author} {\bibfnamefont {P.~A.}\ \bibnamefont {Bushev}}, \bibinfo
  {author} {\bibfnamefont {A.~V.}\ \bibnamefont {Ustinov}},\ and\ \bibinfo
  {author} {\bibfnamefont {M.}~\bibnamefont {Weides}},\ }\bibfield  {title}
  {\bibinfo {title} {Efficient and robust analysis of complex scattering data
  under noise in microwave resonators},\ }\href
  {https://doi.org/10.1063/1.4907935} {\bibfield  {journal} {\bibinfo
  {journal} {Review of Scientific Instruments}\ }\textbf {\bibinfo {volume}
  {86}},\ \bibinfo {pages} {024706} (\bibinfo {year} {2015})},\ \Eprint
  {https://arxiv.org/abs/https://doi.org/10.1063/1.4907935}
  {https://doi.org/10.1063/1.4907935} \BibitemShut {NoStop}%
\bibitem [{\citenamefont {Rieger}\ \emph {et~al.}(2023)\citenamefont {Rieger},
  \citenamefont {G\"unzler}, \citenamefont {Spiecker}, \citenamefont
  {Nambisan}, \citenamefont {Wernsdorfer},\ and\ \citenamefont
  {Pop}}]{Rieger2023}%
  \BibitemOpen
  \bibfield  {author} {\bibinfo {author} {\bibfnamefont {D.}~\bibnamefont
  {Rieger}}, \bibinfo {author} {\bibfnamefont {S.}~\bibnamefont {G\"unzler}},
  \bibinfo {author} {\bibfnamefont {M.}~\bibnamefont {Spiecker}}, \bibinfo
  {author} {\bibfnamefont {A.}~\bibnamefont {Nambisan}}, \bibinfo {author}
  {\bibfnamefont {W.}~\bibnamefont {Wernsdorfer}},\ and\ \bibinfo {author}
  {\bibfnamefont {I.}~\bibnamefont {Pop}},\ }\bibfield  {title} {\bibinfo
  {title} {Fano interference in microwave resonator measurements},\ }\href
  {https://doi.org/10.1103/PhysRevApplied.20.014059} {\bibfield  {journal}
  {\bibinfo  {journal} {Phys. Rev. Appl.}\ }\textbf {\bibinfo {volume} {20}},\
  \bibinfo {pages} {014059} (\bibinfo {year} {2023})}\BibitemShut {NoStop}%
\bibitem [{\citenamefont {Macha}\ \emph {et~al.}(2010)\citenamefont {Macha},
  \citenamefont {van~der Ploeg}, \citenamefont {Oelsner}, \citenamefont
  {Il’ichev}, \citenamefont {Meyer}, \citenamefont {Wünsch},\ and\
  \citenamefont {Siegel}}]{Macha2010}%
  \BibitemOpen
  \bibfield  {author} {\bibinfo {author} {\bibfnamefont {P.}~\bibnamefont
  {Macha}}, \bibinfo {author} {\bibfnamefont {S.~H.~W.}\ \bibnamefont {van~der
  Ploeg}}, \bibinfo {author} {\bibfnamefont {G.}~\bibnamefont {Oelsner}},
  \bibinfo {author} {\bibfnamefont {E.}~\bibnamefont {Il’ichev}}, \bibinfo
  {author} {\bibfnamefont {H.-G.}\ \bibnamefont {Meyer}}, \bibinfo {author}
  {\bibfnamefont {S.}~\bibnamefont {Wünsch}},\ and\ \bibinfo {author}
  {\bibfnamefont {M.}~\bibnamefont {Siegel}},\ }\bibfield  {title} {\bibinfo
  {title} {Losses in coplanar waveguide resonators at millikelvin
  temperatures},\ }\href {https://doi.org/10.1063/1.3309754} {\bibfield
  {journal} {\bibinfo  {journal} {Applied Physics Letters}\ }\textbf {\bibinfo
  {volume} {96}},\ \bibinfo {pages} {062503} (\bibinfo {year} {2010})},\
  \Eprint {https://arxiv.org/abs/https://doi.org/10.1063/1.3309754}
  {https://doi.org/10.1063/1.3309754} \BibitemShut {NoStop}%
\bibitem [{\citenamefont {Sage}\ \emph {et~al.}(2011)\citenamefont {Sage},
  \citenamefont {Bolkhovsky}, \citenamefont {Oliver}, \citenamefont {Turek},\
  and\ \citenamefont {Welander}}]{Sage2011}%
  \BibitemOpen
  \bibfield  {author} {\bibinfo {author} {\bibfnamefont {J.~M.}\ \bibnamefont
  {Sage}}, \bibinfo {author} {\bibfnamefont {V.}~\bibnamefont {Bolkhovsky}},
  \bibinfo {author} {\bibfnamefont {W.~D.}\ \bibnamefont {Oliver}}, \bibinfo
  {author} {\bibfnamefont {B.}~\bibnamefont {Turek}},\ and\ \bibinfo {author}
  {\bibfnamefont {P.~B.}\ \bibnamefont {Welander}},\ }\bibfield  {title}
  {\bibinfo {title} {Study of loss in superconducting coplanar waveguide
  resonators},\ }\href {https://doi.org/10.1063/1.3552890} {\bibfield
  {journal} {\bibinfo  {journal} {Journal of Applied Physics}\ }\textbf
  {\bibinfo {volume} {109}},\ \bibinfo {pages} {063915} (\bibinfo {year}
  {2011})},\ \Eprint {https://arxiv.org/abs/https://doi.org/10.1063/1.3552890}
  {https://doi.org/10.1063/1.3552890} \BibitemShut {NoStop}%
\bibitem [{\citenamefont {Vissers}\ \emph {et~al.}(2012)\citenamefont
  {Vissers}, \citenamefont {Weides}, \citenamefont {Kline}, \citenamefont
  {Sandberg},\ and\ \citenamefont {Pappas}}]{Vissers2012}%
  \BibitemOpen
  \bibfield  {author} {\bibinfo {author} {\bibfnamefont {M.~R.}\ \bibnamefont
  {Vissers}}, \bibinfo {author} {\bibfnamefont {M.~P.}\ \bibnamefont {Weides}},
  \bibinfo {author} {\bibfnamefont {J.~S.}\ \bibnamefont {Kline}}, \bibinfo
  {author} {\bibfnamefont {M.}~\bibnamefont {Sandberg}},\ and\ \bibinfo
  {author} {\bibfnamefont {D.~P.}\ \bibnamefont {Pappas}},\ }\bibfield  {title}
  {\bibinfo {title} {Identifying capacitive and inductive loss in lumped
  element superconducting hybrid titanium nitride/aluminum resonators},\ }\href
  {https://doi.org/10.1063/1.4730389} {\bibfield  {journal} {\bibinfo
  {journal} {Applied Physics Letters}\ }\textbf {\bibinfo {volume} {101}},\
  \bibinfo {pages} {022601} (\bibinfo {year} {2012})},\ \Eprint
  {https://arxiv.org/abs/https://doi.org/10.1063/1.4730389}
  {https://doi.org/10.1063/1.4730389} \BibitemShut {NoStop}%
\bibitem [{\citenamefont {Calusine}\ \emph {et~al.}(2018)\citenamefont
  {Calusine}, \citenamefont {Melville}, \citenamefont {Woods}, \citenamefont
  {Das}, \citenamefont {Stull}, \citenamefont {Bolkhovsky}, \citenamefont
  {Braje}, \citenamefont {Hover}, \citenamefont {Kim}, \citenamefont {Miloshi},
  \citenamefont {Rosenberg}, \citenamefont {Sevi}, \citenamefont {Yoder},
  \citenamefont {Dauler},\ and\ \citenamefont {Oliver}}]{Calusine2018}%
  \BibitemOpen
  \bibfield  {author} {\bibinfo {author} {\bibfnamefont {G.}~\bibnamefont
  {Calusine}}, \bibinfo {author} {\bibfnamefont {A.}~\bibnamefont {Melville}},
  \bibinfo {author} {\bibfnamefont {W.}~\bibnamefont {Woods}}, \bibinfo
  {author} {\bibfnamefont {R.}~\bibnamefont {Das}}, \bibinfo {author}
  {\bibfnamefont {C.}~\bibnamefont {Stull}}, \bibinfo {author} {\bibfnamefont
  {V.}~\bibnamefont {Bolkhovsky}}, \bibinfo {author} {\bibfnamefont
  {D.}~\bibnamefont {Braje}}, \bibinfo {author} {\bibfnamefont
  {D.}~\bibnamefont {Hover}}, \bibinfo {author} {\bibfnamefont {D.~K.}\
  \bibnamefont {Kim}}, \bibinfo {author} {\bibfnamefont {X.}~\bibnamefont
  {Miloshi}}, \bibinfo {author} {\bibfnamefont {D.}~\bibnamefont {Rosenberg}},
  \bibinfo {author} {\bibfnamefont {A.}~\bibnamefont {Sevi}}, \bibinfo {author}
  {\bibfnamefont {J.~L.}\ \bibnamefont {Yoder}}, \bibinfo {author}
  {\bibfnamefont {E.}~\bibnamefont {Dauler}},\ and\ \bibinfo {author}
  {\bibfnamefont {W.~D.}\ \bibnamefont {Oliver}},\ }\bibfield  {title}
  {\bibinfo {title} {Analysis and mitigation of interface losses in trenched
  superconducting coplanar waveguide resonators},\ }\href
  {https://doi.org/10.1063/1.5006888} {\bibfield  {journal} {\bibinfo
  {journal} {Applied Physics Letters}\ }\textbf {\bibinfo {volume} {112}},\
  \bibinfo {pages} {062601} (\bibinfo {year} {2018})},\ \Eprint
  {https://arxiv.org/abs/https://doi.org/10.1063/1.5006888}
  {https://doi.org/10.1063/1.5006888} \BibitemShut {NoStop}%
\bibitem [{\citenamefont {McRae}\ \emph {et~al.}(2018)\citenamefont {McRae},
  \citenamefont {Béjanin}, \citenamefont {Earnest}, \citenamefont {McConkey},
  \citenamefont {Rinehart}, \citenamefont {Deimert}, \citenamefont {Thomas},
  \citenamefont {Wasilewski},\ and\ \citenamefont {Mariantoni}}]{McRae2018}%
  \BibitemOpen
  \bibfield  {author} {\bibinfo {author} {\bibfnamefont {C.~R.~H.}\
  \bibnamefont {McRae}}, \bibinfo {author} {\bibfnamefont {J.~H.}\ \bibnamefont
  {Béjanin}}, \bibinfo {author} {\bibfnamefont {C.~T.}\ \bibnamefont
  {Earnest}}, \bibinfo {author} {\bibfnamefont {T.~G.}\ \bibnamefont
  {McConkey}}, \bibinfo {author} {\bibfnamefont {J.~R.}\ \bibnamefont
  {Rinehart}}, \bibinfo {author} {\bibfnamefont {C.}~\bibnamefont {Deimert}},
  \bibinfo {author} {\bibfnamefont {J.~P.}\ \bibnamefont {Thomas}}, \bibinfo
  {author} {\bibfnamefont {Z.~R.}\ \bibnamefont {Wasilewski}},\ and\ \bibinfo
  {author} {\bibfnamefont {M.}~\bibnamefont {Mariantoni}},\ }\bibfield  {title}
  {\bibinfo {title} {Thin film metrology and microwave loss characterization of
  indium and aluminum/indium superconducting planar resonators},\ }\href
  {https://doi.org/10.1063/1.5020514} {\bibfield  {journal} {\bibinfo
  {journal} {Journal of Applied Physics}\ }\textbf {\bibinfo {volume} {123}},\
  \bibinfo {pages} {205304} (\bibinfo {year} {2018})},\ \Eprint
  {https://arxiv.org/abs/https://doi.org/10.1063/1.5020514}
  {https://doi.org/10.1063/1.5020514} \BibitemShut {NoStop}%
\bibitem [{qki()}]{qkit}%
  \BibitemOpen
  \href@noop {} {\bibinfo {title} {qkit python package}},\ \bibinfo
  {howpublished} {\url{https://github.com/qkitgroup/qkit}}\BibitemShut
  {NoStop}%
\bibitem [{\citenamefont {Ranzani}\ \emph {et~al.}(2013)\citenamefont
  {Ranzani}, \citenamefont {Spietz}, \citenamefont {Popovic},\ and\
  \citenamefont {Aumentado}}]{Ranzani2013}%
  \BibitemOpen
  \bibfield  {author} {\bibinfo {author} {\bibfnamefont {L.}~\bibnamefont
  {Ranzani}}, \bibinfo {author} {\bibfnamefont {L.}~\bibnamefont {Spietz}},
  \bibinfo {author} {\bibfnamefont {Z.}~\bibnamefont {Popovic}},\ and\ \bibinfo
  {author} {\bibfnamefont {J.}~\bibnamefont {Aumentado}},\ }\bibfield  {title}
  {\bibinfo {title} {Two-port microwave calibration at millikelvin
  temperatures},\ }\href {https://doi.org/10.1063/1.4794910} {\bibfield
  {journal} {\bibinfo  {journal} {Review of Scientific Instruments}\ }\textbf
  {\bibinfo {volume} {84}},\ \bibinfo {pages} {034704} (\bibinfo {year}
  {2013})},\ \Eprint {https://arxiv.org/abs/https://doi.org/10.1063/1.4794910}
  {https://doi.org/10.1063/1.4794910} \BibitemShut {NoStop}%
\bibitem [{\citenamefont {Stanley}\ \emph {et~al.}()\citenamefont {Stanley},
  \citenamefont {Parker-Jervis}, \citenamefont {de~Graaf}, \citenamefont
  {Lindström}, \citenamefont {Cunningham},\ and\ \citenamefont
  {Ridler}}]{Stanley2022}%
  \BibitemOpen
  \bibfield  {author} {\bibinfo {author} {\bibfnamefont {M.}~\bibnamefont
  {Stanley}}, \bibinfo {author} {\bibfnamefont {R.}~\bibnamefont
  {Parker-Jervis}}, \bibinfo {author} {\bibfnamefont {S.}~\bibnamefont
  {de~Graaf}}, \bibinfo {author} {\bibfnamefont {T.}~\bibnamefont
  {Lindström}}, \bibinfo {author} {\bibfnamefont {J.~E.}\ \bibnamefont
  {Cunningham}},\ and\ \bibinfo {author} {\bibfnamefont {N.~M.}\ \bibnamefont
  {Ridler}},\ }\bibfield  {title} {\bibinfo {title} {Validating s-parameter
  measurements of rf integrated circuits at milli-kelvin temperatures},\ }\href
  {https://doi.org/https://doi.org/10.1049/ell2.12545} {\bibfield  {journal}
  {\bibinfo  {journal} {Electronics Letters}\ }\textbf {\bibinfo {volume}
  {n/a}}},\ \Eprint
  {https://arxiv.org/abs/https://ietresearch.onlinelibrary.wiley.com/doi/pdf/10.1049/ell2.12545}
  {https://ietresearch.onlinelibrary.wiley.com/doi/pdf/10.1049/ell2.12545}
  \BibitemShut {NoStop}%
\bibitem [{\citenamefont {Johnson}(1928)}]{Johnson1928}%
  \BibitemOpen
  \bibfield  {author} {\bibinfo {author} {\bibfnamefont {J.~B.}\ \bibnamefont
  {Johnson}},\ }\bibfield  {title} {\bibinfo {title} {Thermal agitation of
  electricity in conductors},\ }\href {https://doi.org/10.1103/PhysRev.32.97}
  {\bibfield  {journal} {\bibinfo  {journal} {Phys. Rev.}\ }\textbf {\bibinfo
  {volume} {32}},\ \bibinfo {pages} {97} (\bibinfo {year} {1928})}\BibitemShut
  {NoStop}%
\bibitem [{\citenamefont {Nyquist}(1928)}]{Nyquist1928}%
  \BibitemOpen
  \bibfield  {author} {\bibinfo {author} {\bibfnamefont {H.}~\bibnamefont
  {Nyquist}},\ }\bibfield  {title} {\bibinfo {title} {Thermal agitation of
  electric charge in conductors},\ }\href
  {https://doi.org/10.1103/PhysRev.32.110} {\bibfield  {journal} {\bibinfo
  {journal} {Phys. Rev.}\ }\textbf {\bibinfo {volume} {32}},\ \bibinfo {pages}
  {110} (\bibinfo {year} {1928})}\BibitemShut {NoStop}%
\bibitem [{\citenamefont {Al-Sharadqah}\ and\ \citenamefont
  {Chernov}(2009)}]{Al-Sharadqah2009}%
  \BibitemOpen
  \bibfield  {author} {\bibinfo {author} {\bibfnamefont {A.}~\bibnamefont
  {Al-Sharadqah}}\ and\ \bibinfo {author} {\bibfnamefont {N.}~\bibnamefont
  {Chernov}},\ }\bibfield  {title} {\bibinfo {title} {{Error analysis for
  circle fitting algorithms}},\ }\href {https://doi.org/10.1214/09-EJS419}
  {\bibfield  {journal} {\bibinfo  {journal} {Electronic Journal of
  Statistics}\ }\textbf {\bibinfo {volume} {3}},\ \bibinfo {pages} {886 }
  (\bibinfo {year} {2009})}\BibitemShut {NoStop}%
\bibitem [{\citenamefont {Gao}\ \emph {et~al.}(2007)\citenamefont {Gao},
  \citenamefont {Zmuidzinas}, \citenamefont {Mazin}, \citenamefont {LeDuc},\
  and\ \citenamefont {Day}}]{Gao2007}%
  \BibitemOpen
  \bibfield  {author} {\bibinfo {author} {\bibfnamefont {J.}~\bibnamefont
  {Gao}}, \bibinfo {author} {\bibfnamefont {J.}~\bibnamefont {Zmuidzinas}},
  \bibinfo {author} {\bibfnamefont {B.~A.}\ \bibnamefont {Mazin}}, \bibinfo
  {author} {\bibfnamefont {H.~G.}\ \bibnamefont {LeDuc}},\ and\ \bibinfo
  {author} {\bibfnamefont {P.~K.}\ \bibnamefont {Day}},\ }\bibfield  {title}
  {\bibinfo {title} {Noise properties of superconducting coplanar waveguide
  microwave resonators},\ }\href {https://doi.org/10.1063/1.2711770} {\bibfield
   {journal} {\bibinfo  {journal} {Applied Physics Letters}\ }\textbf {\bibinfo
  {volume} {90}},\ \bibinfo {pages} {102507} (\bibinfo {year} {2007})},\
  \Eprint {https://arxiv.org/abs/https://doi.org/10.1063/1.2711770}
  {https://doi.org/10.1063/1.2711770} \BibitemShut {NoStop}%
\bibitem [{\citenamefont {Barends}\ \emph {et~al.}(2008)\citenamefont
  {Barends}, \citenamefont {Hortensius}, \citenamefont {Zijlstra},
  \citenamefont {Baselmans}, \citenamefont {Yates}, \citenamefont {Gao},\ and\
  \citenamefont {Klapwijk}}]{Barends2008}%
  \BibitemOpen
  \bibfield  {author} {\bibinfo {author} {\bibfnamefont {R.}~\bibnamefont
  {Barends}}, \bibinfo {author} {\bibfnamefont {H.~L.}\ \bibnamefont
  {Hortensius}}, \bibinfo {author} {\bibfnamefont {T.}~\bibnamefont
  {Zijlstra}}, \bibinfo {author} {\bibfnamefont {J.~J.~A.}\ \bibnamefont
  {Baselmans}}, \bibinfo {author} {\bibfnamefont {S.~J.~C.}\ \bibnamefont
  {Yates}}, \bibinfo {author} {\bibfnamefont {J.~R.}\ \bibnamefont {Gao}},\
  and\ \bibinfo {author} {\bibfnamefont {T.~M.}\ \bibnamefont {Klapwijk}},\
  }\bibfield  {title} {\bibinfo {title} {Contribution of dielectrics to
  frequency and noise of nbtin superconducting resonators},\ }\href
  {https://doi.org/10.1063/1.2937837} {\bibfield  {journal} {\bibinfo
  {journal} {Applied Physics Letters}\ }\textbf {\bibinfo {volume} {92}},\
  \bibinfo {pages} {223502} (\bibinfo {year} {2008})},\ \Eprint
  {https://arxiv.org/abs/https://doi.org/10.1063/1.2937837}
  {https://doi.org/10.1063/1.2937837} \BibitemShut {NoStop}%
\bibitem [{\citenamefont {Wolz}\ \emph {et~al.}(2021)\citenamefont {Wolz},
  \citenamefont {McLellan}, \citenamefont {Schneider}, \citenamefont {Stehli},
  \citenamefont {Brehm}, \citenamefont {Rotzinger}, \citenamefont {Ustinov},\
  and\ \citenamefont {Weides}}]{Wolz2021}%
  \BibitemOpen
  \bibfield  {author} {\bibinfo {author} {\bibfnamefont {T.}~\bibnamefont
  {Wolz}}, \bibinfo {author} {\bibfnamefont {L.}~\bibnamefont {McLellan}},
  \bibinfo {author} {\bibfnamefont {A.}~\bibnamefont {Schneider}}, \bibinfo
  {author} {\bibfnamefont {A.}~\bibnamefont {Stehli}}, \bibinfo {author}
  {\bibfnamefont {J.~D.}\ \bibnamefont {Brehm}}, \bibinfo {author}
  {\bibfnamefont {H.}~\bibnamefont {Rotzinger}}, \bibinfo {author}
  {\bibfnamefont {A.~V.}\ \bibnamefont {Ustinov}},\ and\ \bibinfo {author}
  {\bibfnamefont {M.}~\bibnamefont {Weides}},\ }\bibfield  {title} {\bibinfo
  {title} {Frequency fluctuations of ferromagnetic resonances at millikelvin
  temperatures},\ }\href {https://doi.org/10.1063/5.0063668} {\bibfield
  {journal} {\bibinfo  {journal} {Applied Physics Letters}\ }\textbf {\bibinfo
  {volume} {119}},\ \bibinfo {pages} {212403} (\bibinfo {year} {2021})},\
  \Eprint {https://arxiv.org/abs/https://doi.org/10.1063/5.0063668}
  {https://doi.org/10.1063/5.0063668} \BibitemShut {NoStop}%
\bibitem [{\citenamefont {Göppl}\ \emph {et~al.}(2008)\citenamefont {Göppl},
  \citenamefont {Fragner}, \citenamefont {Baur}, \citenamefont {Bianchetti},
  \citenamefont {Filipp}, \citenamefont {Fink}, \citenamefont {Leek},
  \citenamefont {Puebla}, \citenamefont {Steffen},\ and\ \citenamefont
  {Wallraff}}]{Gopple2008}%
  \BibitemOpen
  \bibfield  {author} {\bibinfo {author} {\bibfnamefont {M.}~\bibnamefont
  {Göppl}}, \bibinfo {author} {\bibfnamefont {A.}~\bibnamefont {Fragner}},
  \bibinfo {author} {\bibfnamefont {M.}~\bibnamefont {Baur}}, \bibinfo {author}
  {\bibfnamefont {R.}~\bibnamefont {Bianchetti}}, \bibinfo {author}
  {\bibfnamefont {S.}~\bibnamefont {Filipp}}, \bibinfo {author} {\bibfnamefont
  {J.~M.}\ \bibnamefont {Fink}}, \bibinfo {author} {\bibfnamefont {P.~J.}\
  \bibnamefont {Leek}}, \bibinfo {author} {\bibfnamefont {G.}~\bibnamefont
  {Puebla}}, \bibinfo {author} {\bibfnamefont {L.}~\bibnamefont {Steffen}},\
  and\ \bibinfo {author} {\bibfnamefont {A.}~\bibnamefont {Wallraff}},\
  }\bibfield  {title} {\bibinfo {title} {Coplanar waveguide resonators for
  circuit quantum electrodynamics},\ }\href {https://doi.org/10.1063/1.3010859}
  {\bibfield  {journal} {\bibinfo  {journal} {Journal of Applied Physics}\
  }\textbf {\bibinfo {volume} {104}},\ \bibinfo {pages} {113904} (\bibinfo
  {year} {2008})},\ \Eprint
  {https://arxiv.org/abs/https://doi.org/10.1063/1.3010859}
  {https://doi.org/10.1063/1.3010859} \BibitemShut {NoStop}%
\bibitem [{\citenamefont {Newman}(2005)}]{Newman2005}%
  \BibitemOpen
  \bibfield  {author} {\bibinfo {author} {\bibfnamefont {M.~E.~J.}\
  \bibnamefont {Newman}},\ }\bibfield  {title} {\bibinfo {title} {Power laws,
  pareto distributions and zipf's law},\ }\href
  {https://doi.org/10.1080/00107510500052444} {\bibfield  {journal} {\bibinfo
  {journal} {Contemporary Physics}\ }\textbf {\bibinfo {volume} {46}},\
  \bibinfo {pages} {323} (\bibinfo {year} {2005})},\ \Eprint
  {https://arxiv.org/abs/https://doi.org/10.1080/00107510500052444}
  {https://doi.org/10.1080/00107510500052444} \BibitemShut {NoStop}%
\bibitem [{\citenamefont {Sarsby}(2017)}]{SarsbyThesis}%
  \BibitemOpen
  \bibfield  {author} {\bibinfo {author} {\bibfnamefont {M.}~\bibnamefont
  {Sarsby}},\ }\emph {\bibinfo {title} {Nanoelectronic and nanomechanical
  devices for low temperature applications}},\ \href@noop {} {Ph.D. thesis},\
  \bibinfo  {school} {Lancaster University} (\bibinfo {year}
  {2017})\BibitemShut {NoStop}%
\bibitem [{\citenamefont {Shannon}(1948)}]{Shannon1948}%
  \BibitemOpen
  \bibfield  {author} {\bibinfo {author} {\bibfnamefont {C.~E.}\ \bibnamefont
  {Shannon}},\ }\bibfield  {title} {\bibinfo {title} {A mathematical theory of
  communication},\ }\href {https://doi.org/10.1002/j.1538-7305.1948.tb01338.x}
  {\bibfield  {journal} {\bibinfo  {journal} {The Bell System Technical
  Journal}\ }\textbf {\bibinfo {volume} {27}},\ \bibinfo {pages} {379}
  (\bibinfo {year} {1948})}\BibitemShut {NoStop}%
\bibitem [{\citenamefont {Kraskov}\ \emph {et~al.}(2004)\citenamefont
  {Kraskov}, \citenamefont {St\"ogbauer},\ and\ \citenamefont
  {Grassberger}}]{Kraskov2004}%
  \BibitemOpen
  \bibfield  {author} {\bibinfo {author} {\bibfnamefont {A.}~\bibnamefont
  {Kraskov}}, \bibinfo {author} {\bibfnamefont {H.}~\bibnamefont
  {St\"ogbauer}},\ and\ \bibinfo {author} {\bibfnamefont {P.}~\bibnamefont
  {Grassberger}},\ }\bibfield  {title} {\bibinfo {title} {Estimating mutual
  information},\ }\href {https://doi.org/10.1103/PhysRevE.69.066138} {\bibfield
   {journal} {\bibinfo  {journal} {Phys. Rev. E}\ }\textbf {\bibinfo {volume}
  {69}},\ \bibinfo {pages} {066138} (\bibinfo {year} {2004})}\BibitemShut
  {NoStop}%
\bibitem [{\citenamefont {Noguchi}\ \emph {et~al.}(2019)\citenamefont
  {Noguchi}, \citenamefont {Dominjon}, \citenamefont {Kroug}, \citenamefont
  {Mima},\ and\ \citenamefont {Otani}}]{Noguchi2019}%
  \BibitemOpen
  \bibfield  {author} {\bibinfo {author} {\bibfnamefont {T.}~\bibnamefont
  {Noguchi}}, \bibinfo {author} {\bibfnamefont {A.}~\bibnamefont {Dominjon}},
  \bibinfo {author} {\bibfnamefont {M.}~\bibnamefont {Kroug}}, \bibinfo
  {author} {\bibfnamefont {S.}~\bibnamefont {Mima}},\ and\ \bibinfo {author}
  {\bibfnamefont {C.}~\bibnamefont {Otani}},\ }\bibfield  {title} {\bibinfo
  {title} {Characteristics of very high q nb superconducting resonators for
  microwave kinetic inductance detectors},\ }\href
  {https://doi.org/10.1109/TASC.2019.2904592} {\bibfield  {journal} {\bibinfo
  {journal} {IEEE Transactions on Applied Superconductivity}\ }\textbf
  {\bibinfo {volume} {29}},\ \bibinfo {pages} {1} (\bibinfo {year}
  {2019})}\BibitemShut {NoStop}%
\bibitem [{\citenamefont {Zollitsch}\ \emph {et~al.}(2019)\citenamefont
  {Zollitsch}, \citenamefont {O’Sullivan}, \citenamefont {Kennedy},
  \citenamefont {Dold},\ and\ \citenamefont {Morton}}]{Zollitsch2019}%
  \BibitemOpen
  \bibfield  {author} {\bibinfo {author} {\bibfnamefont {C.~W.}\ \bibnamefont
  {Zollitsch}}, \bibinfo {author} {\bibfnamefont {J.}~\bibnamefont
  {O’Sullivan}}, \bibinfo {author} {\bibfnamefont {O.}~\bibnamefont
  {Kennedy}}, \bibinfo {author} {\bibfnamefont {G.}~\bibnamefont {Dold}},\ and\
  \bibinfo {author} {\bibfnamefont {J.~J.~L.}\ \bibnamefont {Morton}},\
  }\bibfield  {title} {\bibinfo {title} {Tuning high-q superconducting
  resonators by magnetic field reorientation},\ }\href
  {https://doi.org/10.1063/1.5129032} {\bibfield  {journal} {\bibinfo
  {journal} {AIP Advances}\ }\textbf {\bibinfo {volume} {9}},\ \bibinfo {pages}
  {125225} (\bibinfo {year} {2019})},\ \Eprint
  {https://arxiv.org/abs/https://doi.org/10.1063/1.5129032}
  {https://doi.org/10.1063/1.5129032} \BibitemShut {NoStop}%
\bibitem [{\citenamefont {Song}\ \emph {et~al.}(2009)\citenamefont {Song},
  \citenamefont {Heitmann}, \citenamefont {DeFeo}, \citenamefont {Yu},
  \citenamefont {McDermott}, \citenamefont {Neeley}, \citenamefont {Martinis},\
  and\ \citenamefont {Plourde}}]{Song2009}%
  \BibitemOpen
  \bibfield  {author} {\bibinfo {author} {\bibfnamefont {C.}~\bibnamefont
  {Song}}, \bibinfo {author} {\bibfnamefont {T.~W.}\ \bibnamefont {Heitmann}},
  \bibinfo {author} {\bibfnamefont {M.~P.}\ \bibnamefont {DeFeo}}, \bibinfo
  {author} {\bibfnamefont {K.}~\bibnamefont {Yu}}, \bibinfo {author}
  {\bibfnamefont {R.}~\bibnamefont {McDermott}}, \bibinfo {author}
  {\bibfnamefont {M.}~\bibnamefont {Neeley}}, \bibinfo {author} {\bibfnamefont
  {J.~M.}\ \bibnamefont {Martinis}},\ and\ \bibinfo {author} {\bibfnamefont
  {B.~L.~T.}\ \bibnamefont {Plourde}},\ }\bibfield  {title} {\bibinfo {title}
  {Microwave response of vortices in superconducting thin films of re and al},\
  }\href {https://doi.org/10.1103/PhysRevB.79.174512} {\bibfield  {journal}
  {\bibinfo  {journal} {Phys. Rev. B}\ }\textbf {\bibinfo {volume} {79}},\
  \bibinfo {pages} {174512} (\bibinfo {year} {2009})}\BibitemShut {NoStop}%
\bibitem [{\citenamefont {Diener}\ \emph {et~al.}(2012)\citenamefont {Diener},
  \citenamefont {Schellevis},\ and\ \citenamefont {Baselmans}}]{Diener2012}%
  \BibitemOpen
  \bibfield  {author} {\bibinfo {author} {\bibfnamefont {P.}~\bibnamefont
  {Diener}}, \bibinfo {author} {\bibfnamefont {H.}~\bibnamefont {Schellevis}},\
  and\ \bibinfo {author} {\bibfnamefont {J.~J.~A.}\ \bibnamefont {Baselmans}},\
  }\bibfield  {title} {\bibinfo {title} {Homogeneous superconducting phase in
  tin film: A complex impedance study},\ }\href
  {https://doi.org/10.1063/1.4771995} {\bibfield  {journal} {\bibinfo
  {journal} {Applied Physics Letters}\ }\textbf {\bibinfo {volume} {101}},\
  \bibinfo {pages} {252601} (\bibinfo {year} {2012})},\ \Eprint
  {https://arxiv.org/abs/https://doi.org/10.1063/1.4771995}
  {https://doi.org/10.1063/1.4771995} \BibitemShut {NoStop}%
\bibitem [{\citenamefont {Coumou}\ \emph {et~al.}(2013)\citenamefont {Coumou},
  \citenamefont {Driessen}, \citenamefont {Bueno}, \citenamefont {Chapelier},\
  and\ \citenamefont {Klapwijk}}]{Coumou2013a}%
  \BibitemOpen
  \bibfield  {author} {\bibinfo {author} {\bibfnamefont {P.~C. J.~J.}\
  \bibnamefont {Coumou}}, \bibinfo {author} {\bibfnamefont {E.~F.~C.}\
  \bibnamefont {Driessen}}, \bibinfo {author} {\bibfnamefont {J.}~\bibnamefont
  {Bueno}}, \bibinfo {author} {\bibfnamefont {C.}~\bibnamefont {Chapelier}},\
  and\ \bibinfo {author} {\bibfnamefont {T.~M.}\ \bibnamefont {Klapwijk}},\
  }\bibfield  {title} {\bibinfo {title} {Electrodynamic response and local
  tunneling spectroscopy of strongly disordered superconducting tin films},\
  }\href {https://doi.org/10.1103/PhysRevB.88.180505} {\bibfield  {journal}
  {\bibinfo  {journal} {Phys. Rev. B}\ }\textbf {\bibinfo {volume} {88}},\
  \bibinfo {pages} {180505(R)} (\bibinfo {year} {2013})}\BibitemShut {NoStop}%
\bibitem [{\citenamefont {Carter}\ \emph {et~al.}(2019)\citenamefont {Carter},
  \citenamefont {Khaire}, \citenamefont {Chang},\ and\ \citenamefont
  {Novosad}}]{Carter2019}%
  \BibitemOpen
  \bibfield  {author} {\bibinfo {author} {\bibfnamefont {F.~W.}\ \bibnamefont
  {Carter}}, \bibinfo {author} {\bibfnamefont {T.}~\bibnamefont {Khaire}},
  \bibinfo {author} {\bibfnamefont {C.}~\bibnamefont {Chang}},\ and\ \bibinfo
  {author} {\bibfnamefont {V.}~\bibnamefont {Novosad}},\ }\bibfield  {title}
  {\bibinfo {title} {Low-loss single-photon nbn microwave resonators on si},\
  }\href {https://doi.org/10.1063/1.5115276} {\bibfield  {journal} {\bibinfo
  {journal} {Applied Physics Letters}\ }\textbf {\bibinfo {volume} {115}},\
  \bibinfo {pages} {092602} (\bibinfo {year} {2019})},\ \Eprint
  {https://arxiv.org/abs/https://doi.org/10.1063/1.5115276}
  {https://doi.org/10.1063/1.5115276} \BibitemShut {NoStop}%
\bibitem [{\citenamefont {Xu}\ \emph {et~al.}(2019)\citenamefont {Xu},
  \citenamefont {Han}, \citenamefont {Fu}, \citenamefont {Zou},\ and\
  \citenamefont {Tang}}]{Xu2019}%
  \BibitemOpen
  \bibfield  {author} {\bibinfo {author} {\bibfnamefont {M.}~\bibnamefont
  {Xu}}, \bibinfo {author} {\bibfnamefont {X.}~\bibnamefont {Han}}, \bibinfo
  {author} {\bibfnamefont {W.}~\bibnamefont {Fu}}, \bibinfo {author}
  {\bibfnamefont {C.-L.}\ \bibnamefont {Zou}},\ and\ \bibinfo {author}
  {\bibfnamefont {H.~X.}\ \bibnamefont {Tang}},\ }\bibfield  {title} {\bibinfo
  {title} {Frequency-tunable high-q superconducting resonators via wireless
  control of nonlinear kinetic inductance},\ }\href
  {https://doi.org/10.1063/1.5098466} {\bibfield  {journal} {\bibinfo
  {journal} {Applied Physics Letters}\ }\textbf {\bibinfo {volume} {114}},\
  \bibinfo {pages} {192601} (\bibinfo {year} {2019})},\ \Eprint
  {https://arxiv.org/abs/https://doi.org/10.1063/1.5098466}
  {https://doi.org/10.1063/1.5098466} \BibitemShut {NoStop}%
\bibitem [{\citenamefont {Carter}\ \emph {et~al.}(2017)\citenamefont {Carter},
  \citenamefont {Khaire}, \citenamefont {Novosad},\ and\ \citenamefont
  {Chang}}]{Carter2017}%
  \BibitemOpen
  \bibfield  {author} {\bibinfo {author} {\bibfnamefont {F.~W.}\ \bibnamefont
  {Carter}}, \bibinfo {author} {\bibfnamefont {T.~S.}\ \bibnamefont {Khaire}},
  \bibinfo {author} {\bibfnamefont {V.}~\bibnamefont {Novosad}},\ and\ \bibinfo
  {author} {\bibfnamefont {C.~L.}\ \bibnamefont {Chang}},\ }\bibfield  {title}
  {\bibinfo {title} {scraps: An open-source python-based analysis package for
  analyzing and plotting superconducting resonator data},\ }\href
  {https://doi.org/10.1109/TASC.2016.2625767} {\bibfield  {journal} {\bibinfo
  {journal} {IEEE Transactions on Applied Superconductivity}\ }\textbf
  {\bibinfo {volume} {27}},\ \bibinfo {pages} {1} (\bibinfo {year}
  {2017})}\BibitemShut {NoStop}%
\bibitem [{\citenamefont {Shao}\ \emph {et~al.}(2013)\citenamefont {Shao},
  \citenamefont {Jiang}, \citenamefont {Yu}, \citenamefont {Li}, \citenamefont
  {Clements}, \citenamefont {Vollmer}, \citenamefont {Wang}, \citenamefont
  {Xiao},\ and\ \citenamefont {Gong}}]{Shao2013}%
  \BibitemOpen
  \bibfield  {author} {\bibinfo {author} {\bibfnamefont {L.}~\bibnamefont
  {Shao}}, \bibinfo {author} {\bibfnamefont {X.-F.}\ \bibnamefont {Jiang}},
  \bibinfo {author} {\bibfnamefont {X.-C.}\ \bibnamefont {Yu}}, \bibinfo
  {author} {\bibfnamefont {B.-B.}\ \bibnamefont {Li}}, \bibinfo {author}
  {\bibfnamefont {W.~R.}\ \bibnamefont {Clements}}, \bibinfo {author}
  {\bibfnamefont {F.}~\bibnamefont {Vollmer}}, \bibinfo {author} {\bibfnamefont
  {W.}~\bibnamefont {Wang}}, \bibinfo {author} {\bibfnamefont {Y.-F.}\
  \bibnamefont {Xiao}},\ and\ \bibinfo {author} {\bibfnamefont
  {Q.}~\bibnamefont {Gong}},\ }\bibfield  {title} {\bibinfo {title} {Detection
  of single nanoparticles and lentiviruses using microcavity resonance
  broadening},\ }\href {https://doi.org/https://doi.org/10.1002/adma201302572}
  {\bibfield  {journal} {\bibinfo  {journal} {Advanced Materials}\ }\textbf
  {\bibinfo {volume} {25}},\ \bibinfo {pages} {5616} (\bibinfo {year}
  {2013})},\ \Eprint
  {https://arxiv.org/abs/https://onlinelibrary.wiley.com/doi/pdf/10.1002/adma201302572}
  {https://onlinelibrary.wiley.com/doi/pdf/10.1002/adma201302572} \BibitemShut
  {NoStop}%
\bibitem [{\citenamefont {Schneider}(2020)}]{SchneiderThesis}%
  \BibitemOpen
  \bibfield  {author} {\bibinfo {author} {\bibfnamefont {A.}~\bibnamefont
  {Schneider}},\ }\emph {\bibinfo {title} {Quantum Sensing Experiments with
  Superconducting Qubits}},\ \href@noop {} {Ph.D. thesis},\ \bibinfo  {school}
  {Karlsruher Instituts für Technologie} (\bibinfo {year} {2020})\BibitemShut
  {NoStop}%
\bibitem [{\citenamefont {Braumüller}(2018)}]{BraumuellerThesis}%
  \BibitemOpen
  \bibfield  {author} {\bibinfo {author} {\bibfnamefont {J.}~\bibnamefont
  {Braumüller}},\ }\emph {\bibinfo {title} {Quantum simulation experiments
  with superconducting circuits}},\ \href@noop {} {Ph.D. thesis},\ \bibinfo
  {school} {Karlsruher Instituts für Technologie} (\bibinfo {year}
  {2018})\BibitemShut {NoStop}%
\bibitem [{\citenamefont {Jaynes}(1957)}]{Jaynes1957}%
  \BibitemOpen
  \bibfield  {author} {\bibinfo {author} {\bibfnamefont {E.~T.}\ \bibnamefont
  {Jaynes}},\ }\bibfield  {title} {\bibinfo {title} {Information theory and
  statistical mechanics},\ }\href {https://doi.org/10.1103/PhysRev.106.620}
  {\bibfield  {journal} {\bibinfo  {journal} {Phys. Rev.}\ }\textbf {\bibinfo
  {volume} {106}},\ \bibinfo {pages} {620} (\bibinfo {year}
  {1957})}\BibitemShut {NoStop}%
\end{thebibliography}%

\end{document}